\documentclass[superscriptaddress,amsmath,amssymb,aps,pra,twocolumn,floatfix]{revtex4-2}

\usepackage{graphicx}
\usepackage{bm}
\usepackage[colorlinks, linkcolor=mycol, citecolor=mycol, urlcolor=mycol, breaklinks]{hyperref}
\usepackage{braket}
\usepackage{bbold}
\usepackage{bm}
\usepackage{xcolor} 
\usepackage{comment}
\usepackage{physics}

\newcommand{\mi}{ {\rm i} }

\newcommand{\ta}{\textbf{a}}
\newcommand{\tb}{\textbf{b}}
\newcommand{\tc}{\textbf{c}}
\newcommand{\td}{\textbf{d}}
\newcommand{\te}{\textbf{e}}
\newcommand{\tf}{\textbf{f}}

\newcommand{\appname}{Appendix}

\definecolor{mycol}{RGB}{10,55,130}

\begin{document}

\title{Disorder enhanced vibrational entanglement and dynamics in polaritonic chemistry} 

\author{D.~Wellnitz}
\affiliation{IPCMS (UMR 7504), CNRS, 67000 Strasbourg, France}
\affiliation{Universit\'{e} de Strasbourg and CNRS, ISIS (UMR 7006) and icFRC, 67000 Strasbourg, France}

\author{G.~Pupillo}
\affiliation{Universit\'{e} de Strasbourg and CNRS, ISIS (UMR 7006) and icFRC, 67000 Strasbourg, France}

\author{J.~Schachenmayer}
\thanks{schachenmayer@unistra.fr}
\affiliation{IPCMS (UMR 7504), CNRS, 67000 Strasbourg, France}
\affiliation{Universit\'{e} de Strasbourg and CNRS, ISIS (UMR 7006) and icFRC, 67000 Strasbourg, France}

\date{\today}

\begin{abstract}
Collectively coupling molecular ensembles to a cavity has been demonstrated to modify chemical reactions akin to catalysis. Theoretically understanding this experimental finding remains to be an important challenge. In particular the role of quantum effects in such setups is an open question of fundamental and practical interest. Theoretical descriptions often neglect quantum entanglement between nuclear and electro-photonic degrees of freedom, e.g.~by computing Ehrenfest dynamics. Here we discover that disorder can strongly enhance the build-up of this entanglement on short timescales after incoherent photo-excitation. We find that this can have direct consequences for reaction coordinate dynamics. We analyze this phenomenon in a disordered Holstein-Tavis-Cummings model, a minimal toy model that includes all fundamental degrees of freedom. Using a numerical technique based on matrix product states we simulate the exact quantum dynamics of more than 100 molecules. Our results highlight the importance of beyond Born-Oppenheimer theories in polaritonic chemistry.
\end{abstract}

\maketitle

\section{Introduction}

Polaritonic chemistry, or the modification of chemical reactivity using effects of cavity quantum electrodynamics (cavity-QED), is an emerging field of research at the interface of quantum chemistry and physics~\cite{Torma_Strong_2014,Ebbesen_Hybrid_2016,ribeiro2018polariton,flick2018strong,feist2018polaritonic,hertzog2019strong,Herrera2020Mar}. Experiments have demonstrated that a collective coupling of electronic~\cite{Hutchison_Modify_2012,Coles2014,Zhong2016May,Zhong2017Jul,Munkhbat2018Jul,Peters2019Mar,Polak2020,Yu2021May,Mony2021May} or vibrational~\cite{Thomas_Ground_2016,Thomas_Tilting_2018,Vergauwe2019Oct,Lather_Cavity_2019,Hirai2020Sep} transitions of large ensembles of molecules to confined non-local electromagnetic fields can provide means to control chemical reactivity. Many experiments have achieved a collective strong coupling regime, where the cavity and the molecules can coherently exchange energy at a rate faster than their decay processes. In such scenarios, the cavity-molecule system has to be considered as one entity with new ``polaritonic'' eigenstates, which are collective superpositions of photonic and molecular degrees of freedom. Identifying the underlying mechanisms of collective cavity-modified chemistry remains to be a major challenge. A theoretical understanding of the problem requires to solve complex quantum many-body dynamics in large systems with coupled electronic, photonic, and vibrational degrees of freedom.

\medskip

Numerically computing the collective time evolution of all degrees of freedom in polaritonic chemistry is an important --- yet extremely challenging --- task for understanding chemical reaction dynamics, which has been attempted at different levels of approximations. For small systems, the Schr\"odinger equation can be solved directly~\cite{Davidsson2020Atom} or using quantum chemistry tools such as multi-configurational time-dependent Hartree-Fock methods~\cite{vendrell2018coherent}. Density functional theory can be used for ab initio simulations of few realistic molecules~\cite{schafer2019modification}. For larger systems, stronger approximations are needed.Standard approaches are based on the Born-Oppenheimer approximation. In the Born-Oppenheimer approximation, electro-photonic dynamics are treated as instantaneous compared to nuclear dynamics so that polaritonic (and dark) potential energy surfaces can be computed~\cite{galego2015cavity}. On these adiabatic potential energy surfaces, nuclear dynamics can then be computed. However, this method neglects non-adiabatic couplings between potential energy surfaces and thus fails if the separation between potential energy surfaces becomes small, as it is often the case in polaritonic chemistry~\cite{vendrell2018collective,feist2018polaritonic,fabri2021bornoppenheimer}. In order to include the non-adiabatic couplings, two common methods are fewest switches surface hopping~\cite{fregoni2018manipulating,luk2017multiscale,Antoniou2020Oct} or mean-field Ehrenfest dynamics~\cite{groenhof2019tracking,zhang2019non}. Ehrenfest dynamics assumes a product state between nuclear and electro-photonic degrees of freedom, completely neglecting any entanglement between them. As a consequence, such entanglement can serve as a measure for the validity of approximations relying on the separability of nuclear and electro-photonic degrees of freedom, and more generally the complexity of the dynamics.

\smallskip

The role of ``quantum effects'' in molecular dynamics is also a fundamentally interesting research question~\cite{Engel2007Apr,Mohseni2008Nov,caruso2009highly,Collini2010Feb}. Entanglement is often used to determine the importance of quantum effects by quantifying quantum correlations without classical equivalent~\cite{horodecki2009quantum,Amico_Entan_2008,Eisert_Collo_2010}. In this context, the entanglement between electronic and nuclear degrees of freedom of molecules has been previously studied for single molecules~\cite{mckemmish2011quantum,vatasescu2015measures}. For cavity-coupled molecules, it is  known that a collective cavity-coupling can strongly suppress this entanglement by reducing vibronic couplings, an effect termed ``polaron decoupling''~\cite{herrera2016cavity,zeb2018exact}. However this effect neglects local  disorder in the electronic level spacings of individual molecules. In this paper, we will show that the combined effect of local disorder and a cavity coupling can lead to a \textit{strong enhancement} of electro-vibrational entanglement build-up on a typical timescale for coherent molecular dynamics (femtoseconds) after an incoherent photo-excitation.

\smallskip

To analyze this entanglement build-up we make use of a matrix product state (MPS) approach. Recently, MPSs (more broadly: tensor networks) have been suggested to numerically tackle dynamics in polaritonic chemistry~\cite{delPino_Tensor_2018} also for larger system sizes. An MPS can be thought of as a generalization of a product state, which by definition does not include any entanglement, into a larger space with small but finite entanglement. The entanglement of an MPS is limited by a so called ``bond dimension'', which can be systematically increased until convergence is reached~\cite{Schollwock_Density-matrix_2011}. Since excessively large entanglement rarely plays an important role in physical dynamics, MPS simulations often become numerically exact. By construction, MPS concepts provide a direct access for studying the entanglement dynamics of a system, and they have been used in that context extensively, e.g.~for spin-chain or Hubbard-type models in many-body physics~\cite{Eisert_Collo_2010,Amico_Entan_2008}.

\medskip

\begin{figure}
    \centering
    \includegraphics[width=\columnwidth]{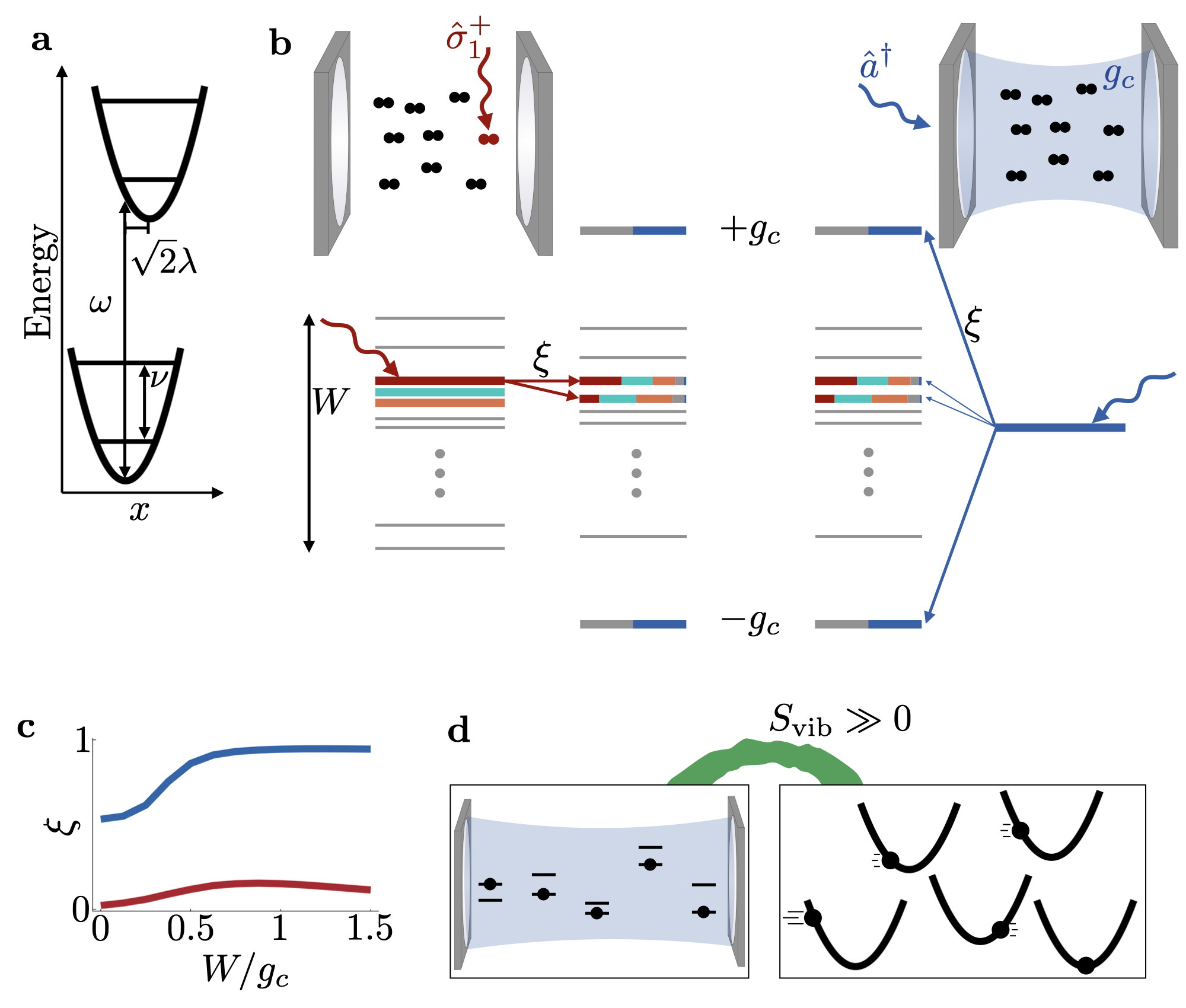}
    \caption{\textbf{Setup and main physics}. \ta{} We consider toy-model molecules with two harmonic potential energy surfaces (vibrational level spacing $\nu$). Both surfaces are energetically separated by the electronic level spacing $\omega$ and displaced by $\sqrt{2}\lambda$ along the reaction coordinate $x$.
    \tb{} An ensemble of molecules is coupled to a cavity with collective strength $g_c$. We analyze dynamics after incoherent photo-excitation of either an individual molecule (red, left) or the cavity (blue, right). An energy level scheme for electro-photonic excitations is sketched. Disorder leads to inhomogeneous broadening by $W$ (left). The coupling of $N$ electronic excitation states (left; gray, red, cyan, orange lines) and a single cavity excitation (right; blue line) lead to new eigenstates (center) that are superpositions with contributions indicated by the different colors. For $g_c\gg W$, two polariton states at energies $\pm g_c$ are formed (half grey, half blue), as well as $N-1$ dark states (other lines). Due to disorder, the dark states are superpositions of a few energetically resonant electronic excitations. All dark states also acquire a small photon weight (very small blue contribution). After incoherent excitation, energy is transferred through the coupled eigenstates as indicated by straight arrows (transfer probability $\xi$). For a molecule excitation, energy is predominantly transferred through dark states, for a cavity excitation through polariton states (arrow thickness).
    \tc{} Disorder enhances the transfer away from the initially excited state  after molecular (red) or cavity (blue) excitation. The plot shows a time averaged transfer probability $\xi = \nu/(2\pi) \int_0^{2\pi/\nu} dt \, [1 - \langle\hat O^\dagger \hat O\rangle(t)]$ where $\hat O =  \hat \sigma_1^-, \hat a$ in a system with 100 molecules.
    \td{} Excitation transfer leads to a coherent out-of-phase oscillation of different molecules and thus large entanglement entropy, $S_\mathrm{vib}$, between electro-photonic (left) and vibrational (right) degrees of freedom.}
    \label{fig:fig1}
\end{figure}

Here, using this numerical approach we study the femtosecond-scale dynamics of more than 100 molecules with electronic transitions collectively strongly coupled to a cavity mode (electronic strong coupling) after an incoherent photo-excitation (see Fig.~\ref{fig:fig1} for a sketch). We analyze a minimal disordered version of the Holstein-Tavis-Cummings (HTC) model~\cite{cwik2014polariton,herrera2016cavity,herrera2018theory}, which despite its simplicity includes the main ingredients for microscopically understanding physical mechanisms in polaritonic chemistry. We find that disorder enhances excitation transfer from the initially excited state to a number of molecules selected by a resonance condition [see Fig.~\ref{fig:fig1}(b/c)]~\cite{botzung2020dark,chavez2021disorder,Dubail2021}. This leads to coherent out-of-phase oscillations of the vibrational modes of these molecules. As a consequence, disorder enhances entanglement between vibrations and electronic degrees of freedom several-fold [see sketch in Fig.~\ref{fig:fig1}(d)]. This effect is largest in a regime where disorder is energetically comparable to collective cavity-couplings. Importantly, we find that the disorder-induced focused excitation transfer to a few molecules leads to an enhanced cavity-modified vibrational dynamics on the single molecule level, compared to a disorder-less scenario where the excitation is diluted among all coupled molecules equally. This effect crucially depends on whether the initial incoherent excitation is absorbed by a single molecule or the cavity, and we analyze both scenarios [see Fig.~\ref{fig:fig1}(b)]. We further relate large entanglement to modifications of the shape of the nuclear wave packets, which become broadened and non-Gaussian. In this respect, the vibrational entanglement may have direct consequences for chemical processes.

\section{Results}
{\textbf{Theoretical Model} -- }
We consider a system of $N$ toy model molecules coupled to a single mode optical cavity, i.e.~a disordered version of the Holstein-Tavis-Cummings (HTC) model~\cite{cwik2014polariton,herrera2016cavity,herrera2018theory}. Here, each molecule has two electronic energy levels. Different nuclear equilibrium configurations in the ground and excited state result in two displaced harmonic one dimensional potential energy ``surfaces'' as shown in Fig.~\ref{fig:fig1}a. We further include an inhomogeneous broadening, i.e.~disorder of the electronic energy stemming from random energy spacings of the electronic levels~\cite{Houdre_Vacuum_1996}, typically induced by the environment in experiments. The disordered HTC Hamiltonian reads~\cite{herrera2016cavity}

\begin{align}
    \hat H &= \hat H_\mathrm{TC} + \hat H_\mathrm{vib} + \hat H_\mathrm{H} + \hat H_\mathrm{dis} \, . \label{eq:ham}
\end{align}
The coupling of the cavity is described by the Tavis-Cummings (TC) Hamiltonian, which, in a frame rotating at the cavity frequency $\omega_\mathrm{C}$, reads ($\hbar = 1$ throughout this paper)

\begin{align}
    \hat H_\mathrm{TC} &= \sum_{n=1}^N \Delta \hat \sigma_n^+ \hat \sigma_n^- + g \sum_{n=1}^N \qty(\hat a \hat \sigma_n^+ + \hat a^\dagger \hat \sigma_n^-), \label{eq:ham-tc}
\end{align}
where $\hat a$ is the destruction operator for a cavity photon, $\hat \sigma^\pm_n$ are the raising/lowering operators for the electronic level of the $n$-th molecule. $\Delta = \omega - \omega_\mathrm{C}$ is the detuning between the electronic transition frequency at the Condon point $\omega$ and $\omega_\mathrm{C}$, chosen to be $\Delta=0$ in the remainder of this paper. The coupling strength of a single molecule to the cavity is given by $g \equiv g_c/\sqrt{N}$. In the single-excitation Hilbert space considered here, the TC Hamiltonian has two polariton eigenstates $\ket{\pm} = \hat a^\dagger / \sqrt{2} \pm \sum_n \hat \sigma_n^+ / \sqrt{2N}\ket{0}_\mathrm{exc+ph}$ for the ground state $\ket{0}_\mathrm{exc+ph}$ without any excitations, split by the Rabi splitting of $2g_c$. The other $N-1$ eigenstates are degenerate dark states with zero energy.

The nuclear coordinates are described by harmonic potentials

\begin{align}
    \hat H_\mathrm{vib} &= \nu \sum_{n=1}^N \hat b^\dagger_n \hat b_n \, , \label{eq:ham-vib}
\end{align}
where $\hat b_n$ is the lowering operator of the $n$-th molecule and $\nu$ the molecular oscillation frequency. The eigenstates of $\hat H_\mathrm{vib}$ are Fock states $\prod_n (\hat b_n^\dagger)^{a_n} \ket{0_\mathrm{vib}}$ with $a_n$ vibrational quanta on the $n$-th molecule and the total (undisplaced) vibrational ground state $\ket{0_\mathrm{vib}}$. We define dimensionless oscillator position and momentum variables as $\hat x_n = (\hat b_n + \hat b_n^\dagger)/\sqrt{2}$ and $\hat p_n = -\mi (\hat b_n - \hat b_n^\dagger) / \sqrt{2}$, respectively.

The nuclear coordinate of each molecule is coupled to its electronic state by a Holstein coupling

\begin{align}
    \hat H_\mathrm{H} &= -\lambda \nu \sum_{n=1}^N \qty(\hat b_n + \hat b_n^\dagger) \hat \sigma_n^+ \hat \sigma_n^- \, . \label{eq:ham-h}
\end{align}
This corresponds to a shift of the excited state potential energy surface. The dimensionless Huang-Rhys factor $\lambda^2$ quantifies the minimum of the excited state harmonic potential at position $\sqrt{2}\lambda$ with energy $\omega - \lambda^2\nu$.

Finally, we include disorder by

\begin{align}
    \hat H_\mathrm{dis} = \sum_n \epsilon_n \hat \sigma_n^+\hat \sigma_n^- \, ,
\end{align}
where $\epsilon_n = \omega_n - \omega$ is the deviation of the electronic transition energy of the $n$-th molecule from the mean. We take the $\epsilon_n$ as independent, normally distributed random variables with mean 0 and variance $W^2$.

\medskip

{\textbf{Dynamics \& Entanglement} -- }  In the following, we analyze the short-time Hamiltonian dynamics on the scale of a single nuclear vibration period, $0\leq t \leq 2\pi/\nu$ for two different initial states. In one case, a single molecule ($n=1$) is excited by the incoherent absorption of a photon, i.e.~we consider the initial state $\ket{\psi_0^m}  = \hat \sigma^+_1 \ket{0}_{\rm ph}  \ket{0}_{\rm exc} \ket{0}_{\rm vib}$ [Fig.~\ref{fig:fig1}(b), left]. In the other case the photon is incoherently absorbed by the cavity, $\ket{\psi_0^c}  = \hat a^\dag  \ket{0}_{\rm ph}  \ket{0}_{\rm exc}  \ket{0}_{\rm vib}$ [Fig.~\ref{fig:fig1}(b), right]. Here, $\ket{0}_{\rm exc,vib,ph}$ denote the respective ground states of the bare electronic, vibrational and photonic Hamiltonian.

In order to analyze entanglement between electro-photonic and nuclear degrees of freedom, we separate the full Hilbert space as $\mathcal H = \mathcal H_\mathrm{ph} \otimes \mathcal H_\mathrm{exc} \otimes \mathcal H_\mathrm{vib}$ into three sub-Hilbert spaces for the cavity photon, electronic excitations, and vibrations, respectively. For a pure state $\ket \psi$, the entanglement between the two subsystems $\mathcal H_\mathrm{ph} \otimes \mathcal H_\mathrm{exc} $ and $ \mathcal H_\mathrm{vib}$ can be quantified by the von Neumann entropy of either subsystem~\cite{Amico_Entan_2008,Eisert_Collo_2010},~e.g.

\begin{align}
    S_\mathrm{vib} = - \Tr \left[\hat \rho_\mathrm{vib}\log_2(\hat \rho_\mathrm{vib})\right],
\end{align}
where $\hat \rho_\mathrm{vib}$ is the reduced density matrix which can be obtained from the state $\ket \psi$ by tracing over $\mathcal H_\mathrm{ph} \otimes \mathcal H_\mathrm{exc}$: $\hat \rho_\mathrm{vib} = \Tr_{\mathrm{ph}+\mathrm{exc}} (\ket{\psi}\bra{\psi})$. In the case of a product state (or ``mean-field'') assumption, the state of the system would be assumed to factorize throughout the evolution of the system

\begin{align}
\label{eq:ps_ansatz}
\ket{\psi(t)} = \ket{\phi_{\rm ph+exc}(t)} \otimes \ket{\phi_{\rm vib}(t)}.
\end{align}
In this scenario, $\hat \rho_{\rm vib} (t) = \ket{\phi_{\rm vib}(t)}\bra{\phi_{\rm vib}(t)}$ and $S_{\rm vib}(t) = 0$ at all times. An entangled state $\ket \psi$ is a linear superposition of many such terms, resulting in $S_\mathrm{vib} > 0$. The von Neumann entropy $S_\mathrm{vib}$ can be readily computed in the MPS framework (see \appname{}). It is noteworthy that this product state assumption is equivalent to the one made in mean-field Ehrenfest dynamics, where in addition the nuclear motion is treated classically. Since in our case the nuclear wavefunction always stays coherent and thus follows classical equations of motion, our product state results are equivalent to mean-field Ehrenfest results.

\medskip

{\textbf{Parameter regimes} -- } We choose parameter values that are motivated by a setup with Rhodamine 800, for which strong coupling has been demonstrated~\cite{valmorra2011strong}, and which has been previously considered in tensor network studies of strong coupling experiments~\cite{delPino_Tensor_2018}. In particular, we set $0.1 \leq \lambda \leq 0.5$ and $\nu=0.3g_c$. For an experimentally demonstrated vacuum Rabi splitting of $2g_c = 700$meV~\cite{Hutchison_Modify_2012}, this corresponds to $\nu = 105$meV and reorganization energies $1\textrm{meV} \lesssim \lambda^2\nu \lesssim 26$meV, similar to measured values~\cite{christensson2010electronic}. Thermal excitation fractions $\sim \exp(-\nu / k_BT)$ are negligible at room temperature ($k_BT \approx 26$meV). Although the Rabi splitting of $700$meV falls into the ultra-strong coupling regime for the relevant electronic transition of Rhodamine 800 at $\sim 2$eV~\cite{valmorra2011strong}, we do not include counter-rotating terms here in order to derive general results which are relevant for strong coupling experiments.

For our case of $\lambda\nu \ll \nu \ll g_c$, the Hamiltonian Eq.~\eqref{eq:ham} can be categorized into strong ($W\ll g_c)$ and weak ($W\gg g_c$) coupling regimes depending on the relative magnitude of $\hat H_\mathrm{TC}$ and $\hat H_\mathrm{dis}$. The strong coupling regime features polaritonic and dark eigenstates of $\hat H_\mathrm{TC}$ which are mixed perturbatively (Fig.~\ref{fig:fig1}b). In perturbation theory we find that ``gray'' states $\ket d$ acquire photo-contributions of $\sum_d \abs{\bra{d}\ket{1_\mathrm{ph}}}^2 \sim \lambda^2\nu^2/(2g_c^2)$ and $\sum_d \abs{\bra{d}\ket{1_\mathrm{ph}}}^2 \approx W^2/g_c^2$ due to small vibronic coupling and disorder, respectively (see \appname{} and~\cite{Houdre_Vacuum_1996,agranovich2003cavity,michetti2009exciton,Dubail2021} for details). In the weak coupling regime, polariton states cease to exist and all eigenstates are structurally similar to the ``gray'' states in Fig.~\ref{fig:fig1}a. We vary $0 \leq W \leq 1.5g_c$ analyzing both  weak and strong coupling scenarios. The timescale of vibrational evolution $t \sim 2\pi / \nu$ corresponds to tens of femtoseconds, and can be faster than dissipative mechanisms which we do not include explicitly. For quality factors $Q \gtrsim 1000$ which have e.g.~been achieved for distributed Bragg reflectors, cavity decay is negligible on these timescales~\cite{hou2020ultralong}. Similarly, relaxation of molecular excitation into vibrational or electromagnetic reservoirs typically occurs on even slower timescales of picoseconds or nanoseconds, respectively~\cite{herrera2018theory}. In fact, on a microscopic level the coherent dynamics due to disorder and vibronic coupling terms $\hat H_\mathrm{dis}$ and $\hat H_\mathrm{H}$ that we simulate here can be considered as one of the mechanisms responsible for electronic dephasing.

\begin{figure}
    \centering
    \includegraphics[width=\columnwidth]{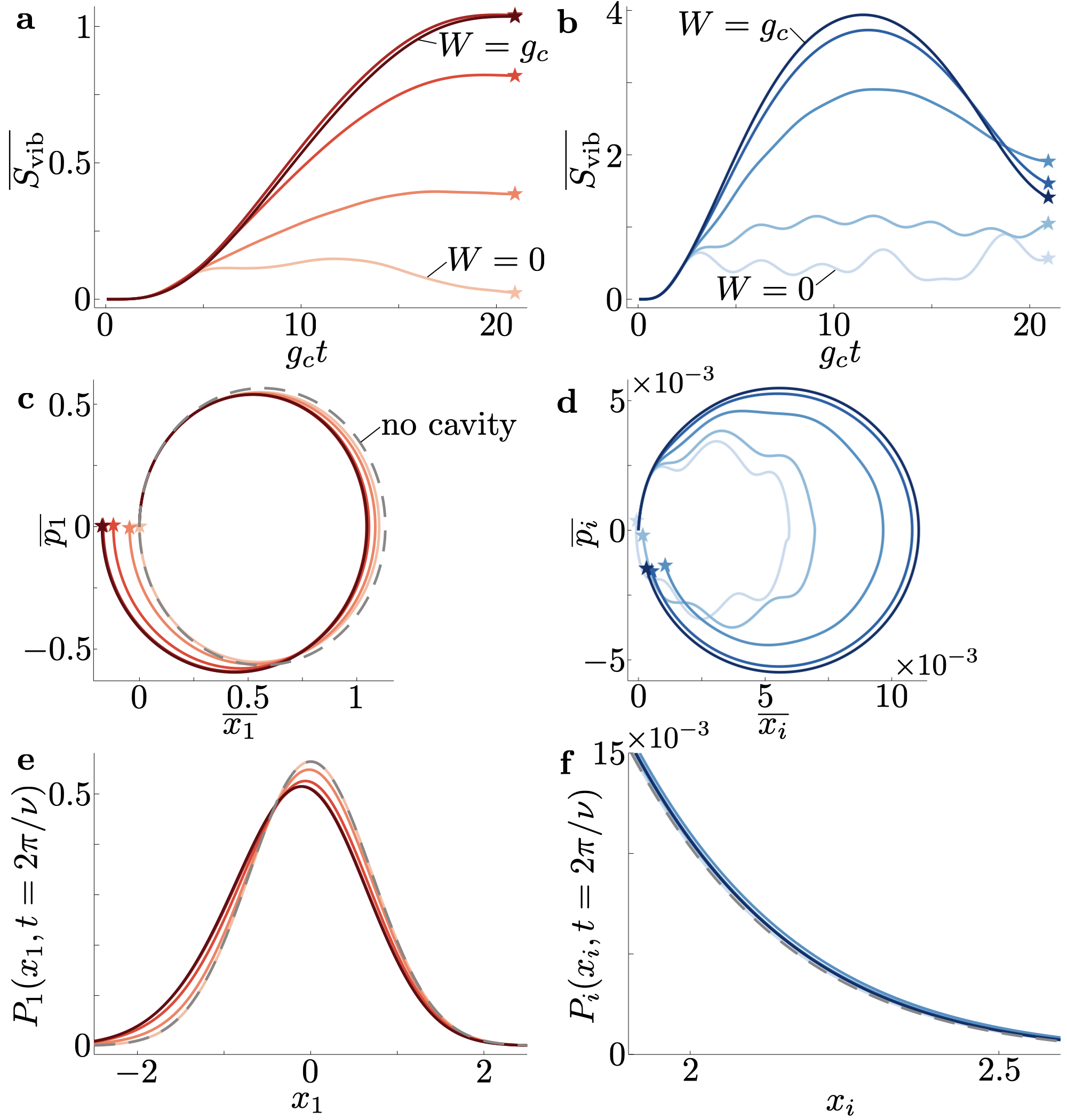}
    \caption{\textbf{Main results}. \ta{},\tb{}~Time evolution of the disorder averaged entanglement entropy $\overline{S_\mathrm{vib}}$ in the time-range $0 \leq t \leq 2\pi/\nu$ for disorder strengths $0 \leq W \leq g_c$ (from light to dark $W=0$, $g_c/4$, $g_c/2$, $3g_c/4$, $g_c$). Stars indicate the final time $t=2\pi/\nu$. The left panels (red lines) correspond to the initial molecule excitation state $\ket{\psi_0^m}$, the right panels (blue lines) to the initial cavity excitation $\ket{\psi_0^c}$. \tc{},\td{} Vibrational phase space evolution. Shown are the disorder averaged expectation values $\overline{x_i}$ and $\overline{p_i}$ of the oscillator position and momentum operators, $\hat x_i$ and $\hat p_i$, for values of $W$ correpsonding to \ta{}. The gray dashed-line shows the no-cavity case.  In \tc{} the dynamics of the initially excited molecule is shown, in \td{}{} an additional average over all molecules is taken. \te{},\tf{} Averaged probability distributions of the reaction coordinate $x$ at time $t=2\pi/\nu$ (stars in other panels, gray dashed line: no-cavity case). In all panels, we average over 64 disorder realizations, $N=100$, $\nu = 0.3g_c$, $\lambda=0.4$.}
    \label{fig:distributions}
\end{figure}

\medskip

{\textbf{Main results} -- }
Fig.~\ref{fig:distributions} visualizes the main feature of the entanglement and vibrational dynamics after initial molecular ($\ket{\psi_0^m}$, panels a,c,e), and cavity ($\ket{\psi_0^c}$, panels b,d,f) excitation. Strikingly, in both scenarios we find that increasing the disorder in the range $0 \leq W \leq g_c$ leads to a drastically enhanced entanglement entropy build-up, seen in the evolution of the disorder averaged entropy $\overline{S_{\rm vib}}$ in Fig.~\ref{fig:distributions}a,b. For $W=0$, the entanglement entropy remains below values of one, and, in the cavity excitation case only, exhibits oscillatory features which we can attribute to collective Rabi oscillation due to a predominant excitation transfer to the polariton states. For $W>g_c/2$ those features disappear and we observe  a strong increase to a maximum value at $t \sim 2\pi/\nu$ and $t\sim \pi/\nu$ in Fig.~\ref{fig:distributions}a and b, respectively. This entanglement build-up is matched by modifications of the phase space dynamics (Fig.~\ref{fig:distributions}c,d) and the shape of the probability distribution $P_i(x_i,t)$ (Fig.~\ref{fig:distributions}e,f) of the reaction coordinate. Below, we will relate all three effects to disorder enhanced excitation transfer. We will see that neither the entanglement build-up nor the distribution shape changes can be captured by a product state assumption Eq.~\eqref{eq:ps_ansatz}, i.e.~they go beyond the mean-field Ehrenfest dynamics.

\medskip{}

Fig.~\ref{fig:distributions}c,d shows the phase space dynamics of the disorder averaged expectation values $\overline{x_i}$ and $\overline{p_i}$ of the reaction coordinate position and momentum operators $\hat x_i$ and $\hat p_i$ on molecule $i$, respectively. In the molecular excitation case in Fig.~\ref{fig:distributions}c, we observe phase space circles for the initially excited molecule ($i=1$). In the disorder-less case $W=0$, we find an oscillation around the displaced equilibrium position of the excited state oscillator, $\sqrt{2} \lambda \approx 0.57$. In this case the evolution is very close to the no-cavity scenario (gray dashed line), for which we obtain a perfect circle around $\sqrt{2} \lambda$ corresponding to the usual coherent harmonic oscillator evolution. 

However, the situation changes drastically for $W>0$. Now, the centers of the phase space circles dynamically shift to smaller values of $\overline{x_1}$. We note that this behavior can be rationalized without requiring the large entanglement build-up seen in Fig.~\ref{fig:distributions}a. Assuming the product state ansatz from Eq.~\eqref{eq:ps_ansatz}, one would expect that the Holstein term $\hat H_\mathrm{H}$ [Eq.~\eqref{eq:ham-h}] leads to an effective excited state oscillator equilibrium position of $\sqrt{2} \lambda \langle \hat \sigma_1^+\hat \sigma_1^- \rangle$ and thus effectively to a time-dependent shift of the minimum depending on $\langle \sigma_1^+\hat \sigma_1^- \rangle (t)$. For $W=0$ the initial state $\ket{\psi_0^m}$ is almost a dark eigenstate of $\hat H_\mathrm{TC}$, such that cavity induced excitation transfer is strongly suppressed and the excitation remains on the molecule, $\langle \hat \sigma_1^+\hat \sigma_1^- \rangle (t) \sim 1$~(see \appname{}). In contrast, for finite disorder $W>0$, the excitation transfer is significantly enhanced and we perturbatively derive $1 - \langle \hat \sigma_1^+ \hat \sigma_1^- \rangle (t) \sim Wt/N$ for $g \ll W \ll g_c$~(see \appname{}). This is in qualitative agreement with recent results predicting that disorder can enhance excitation transfer in models without vibrations~\cite{botzung2020dark, chavez2021disorder, Dubail2021}.

For an initial cavity excitation $\ket{\psi_0^c}$, the phase space evolution traces much smaller circles. As expected, for $W=0$ the phase space evolution is approximately centered at $\overline{x_i} \sim \sqrt{2}\lambda/(2N)$, and exhibits oscillations at polariton Rabi frequencies. For increasing disorder $W \rightarrow g_c$, the center of the circle now shifts in the opposite direction compared to Fig.~\ref{fig:distributions}c, to roughly twice the value $\overline{x_i} \rightarrow \sqrt{2}\lambda/N$. This can again be rationalized by looking at the evolution of the expected local molecule excitations $\langle\hat \sigma_i^+\hat \sigma_i^- \rangle(t)$. For $W = \lambda\nu = 0$, the hybrid nature of the polariton states induces Rabi oscillations between the initial cavity photon state and a collective excitation of all molecules, such that for each molecule the excitation fraction oscillates according to $\langle \hat \sigma_i^+\hat \sigma_i^- \rangle(t) = \cos^2(g_ct)/N$, leading to the observed phase space evolution for $W=0$. Finite disorder, however, leads to a photo-contribution of all dark states (perturbatively $\sim W^2/g_c^2$). Therefore, excitations are now transferred quickly (timescale $1/g_c$) from the cavity to individual molecules, and one thus expects a disorder averaged excitation population on each molecule $\overline{\langle \hat \sigma_i^+\hat \sigma_i^- \rangle} \to 1/N$ for sufficiently large $W$. This explains the observed shift.

\medskip

Fig.~\ref{fig:distributions}e,f shows the probability distribution $P_i(x_i,t)$ of the reaction coordinate at time $t=2\pi/\nu$. Without cavity, at this time the distribution is a Gaussian centered at $x_1 = 0$ with variance $1/2$, corresponding to a coherent state (grey dashed line). We find that in a cavity and for $W=0$, the distribution is extremely close to the no-cavity scenario. For increasing $W$, however, for the molecular excitation $\ket{\psi_0^m}$, the distribution of the reaction coordinate of the initially excited molecule $P_1(x_1, t=2\pi/\nu)$ clearly shifts to smaller values of $x_1$. In addition, the distribution broadens and acquires an asymmetric shape in Fig.~\ref{fig:distributions}e. For an initial cavity excitation $\ket{\psi_0^c}$, we observe that finite $W$ leads to modifications in the tails of the distribution only, i.e.~for large values of $x_i$ (Fig.~\ref{fig:distributions}f). Note that the tail modifications in Fig.~\ref{fig:distributions}f seem very small, since a single molecule only receives a $\sim 1/N$ contribution of the excitation energy (here, $N=100$). However, below we see that the cumulative effect on the wave function shape can still have important consequences for many molecules.

\begin{figure}
    \centering
    \includegraphics[width=\columnwidth]{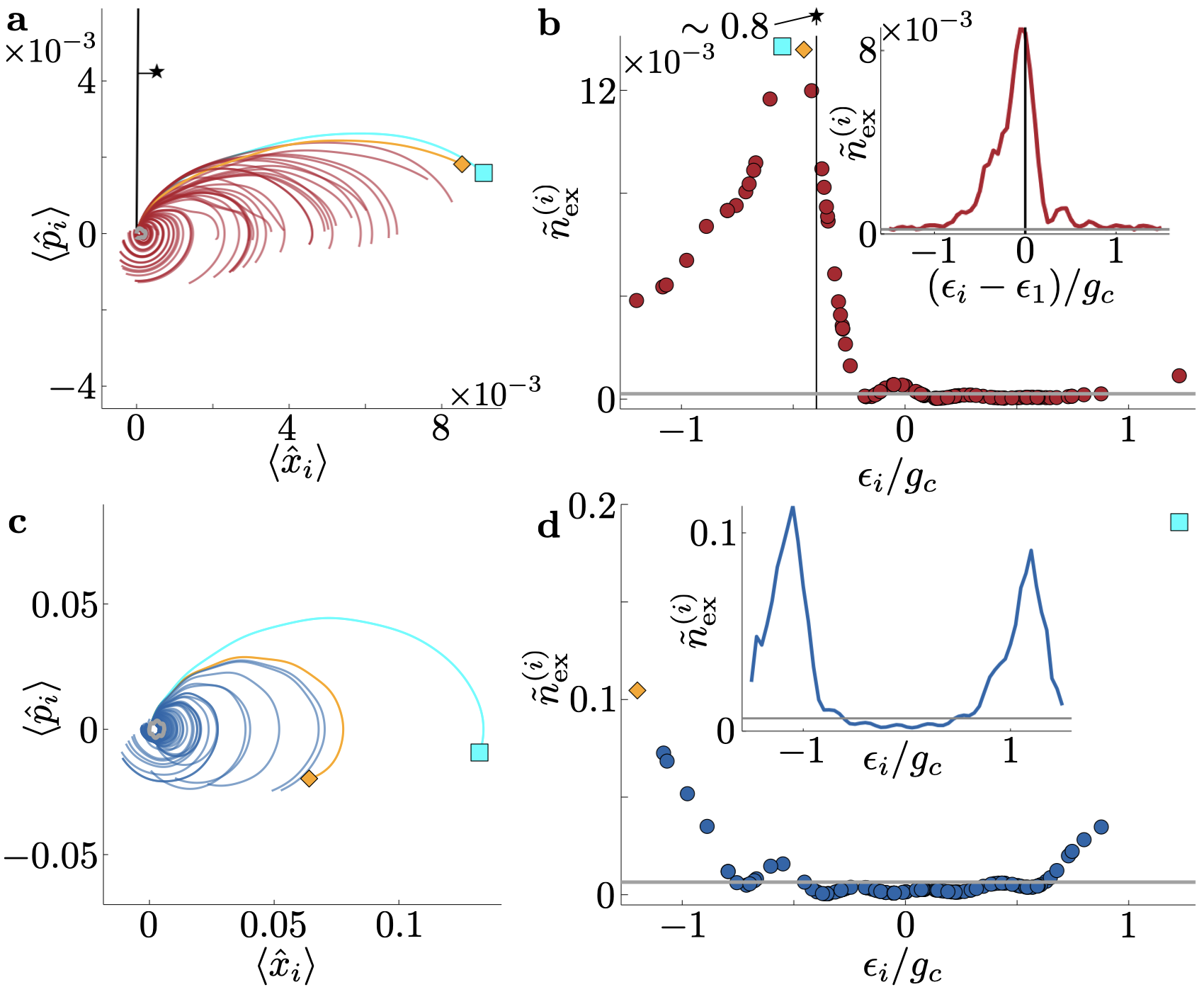}
    \caption{\textbf{Excitation transfer dynamics.} \ta{} Microscopic phase space evolution of 100 molecular oscillators  for a single disorder realization with $W=g_c/2$. The black star and line correspond to the initially excited molecule, the other lines to the $99$ initially un-excited ones. The gray line (barely visible around the origin) represents the $W=0$ reference. The cyan square and orange diamond are two example molecules with most strongly modified vibrational dynamics, also identified in \tb{}.  \tb{} Excitation probability $\tilde n_\mathrm{ex}^{(i)} = \langle \hat \sigma_i^+ \hat \sigma_i^- \rangle (t=2\pi/\nu)$ as a function of the energy off-set $\epsilon_i$ of the respective molecule. The inset shows the disorder-averaged excitation probability as a function of the energy difference to the initially excited state. The gray horizontal lines are the $W=0$ reference. \tc{},\td{} identical plots for a cavity excitation. Parameters: $N=100$, $\nu = 0.3g_c$, $\lambda=0.4$, disorder drawn from a normal distribution with width $W$, 256 disorder realizations.}
    \label{fig:fluctuations}
\end{figure}

\medskip

{\textbf{Excitation transfer dynamics} -- }In Fig.~\ref{fig:fluctuations}, we now exemplify the connection between the time-dependent local molecular excitation and the phase space evolution for an intermediate disorder strength $W = g_c/2$ microscopically. Fig.~\ref{fig:fluctuations}a shows the vibrational evolution of each of the $100$ molecules for a single disorder realization, after exciting one molecule initially (line with star: excited molecule, other red/cyan/yellow lines: $99$ initially unexcited molecules). Strikingly, we observe that the molecules whose dynamics is modified most strongly correspond to the ones with a random energy very close to the initially excited one. The reason for this is seen in comparison with Fig.~\ref{fig:fluctuations}b where we plot the excitation numbers of the molecules at $t=2\pi/\nu$ as function of their random energy offset $\epsilon_i$. There we identify the molecules with the strongest phase space modification (cyan square and orange diamond), and the initially excited one (blue star and vertical line). Crucially, the excitation fraction of these molecules, and thus their phase space dynamics, is much larger than for the homogeneous disorder-less case with $W=0$ (gray lines in Fig.~\ref{fig:fluctuations}a,b, barely visible in a). The same behavior is generally seen also after disorder-averaging (see inset). We attribute a visible asymmetry towards smaller energies in Fig.~\ref{fig:fluctuations}b to additional resonances with states of higher vibrational energies. It is also interesting to point out that in contrast to the initially excited molecule, the phase space variables of the other molecules generally do not complete one revolution until $t=2\pi/\nu$ (Fig.~\ref{fig:fluctuations}a), and all molecular oscillators evolve out-of-phase.

\medskip

A similar picture presents itself when initially exciting the cavity-mode (Fig.~\ref{fig:fluctuations}c,d). The excitation is again primarily transferred from the cavity to several molecules, but now with energies $\epsilon_i \sim \pm g_c$ close to resonance with the bare polaritons in the strong coupling regime. These molecules acquire much larger excitation fractions than the no disorder reference (gray line in Fig.~\ref{fig:fluctuations}d). As a result, molecular oscillations of molecules with an energy offset $\epsilon_n \sim \pm g_c$ are most strongly modified, as confirmed in Fig.~\ref{fig:fluctuations}c.

\medskip

We can deduce the following microscopic picture from our analysis in Fig.~\ref{fig:fluctuations}: While in the $W=0$ case the initial excitation is generally diluted throughout the system, disorder $W>0$ leads to a strongly enhanced excitation transfer to a few molecules in the energetic vicinity of either the initially excited molecule or the polariton states, depending on the scenario (as sketched in Fig.~\ref{fig:fig1}b). In a product state picture, this then modifies the vibrational dynamics of those molecules depending on the amount of local excitation, $\langle \hat \sigma_i^+ \hat \sigma_i^- \rangle$. However, the product state assumption contradicts the build-up of large vibrational entanglement seen in Fig.~\ref{fig:distributions}a/b. Rather, the  out-of-phase oscillator dynamics should be considered quantum-mechanically coherent, leading to the large entanglement entropies. In the following we will study the direct implications of this entanglement.

\medskip

\begin{figure}
    \centering
    \includegraphics[width=\columnwidth]{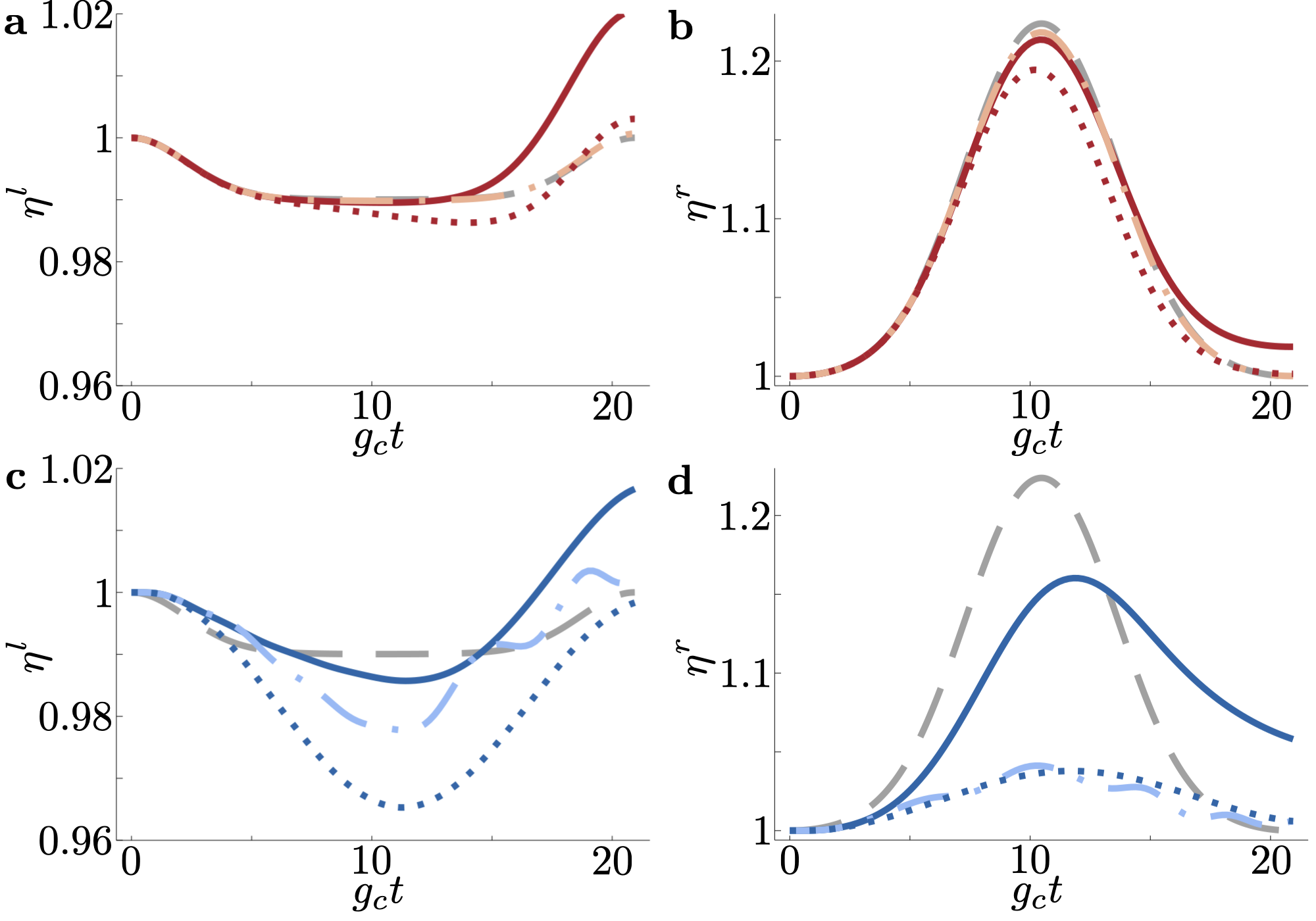}
    \caption{\textbf{Tail weight dynamics.} \ta{},\tb{} Time evolution of the cumulative left tail weight $\eta^l$ (\ta{}) and the right tail weight $\eta^r$ (\tb{}) as defined in Eq.~\eqref{eq:def_tail} for an initial molecule excitation. The dark red solid line is the disorder-averaged exact time evolution for $W=g_c/2$, whereas the red dotted line shows equivalent results computed with a product state approximation [Eq.~\eqref{eq:ps_ansatz}]. The light red dash-dotted line displays the results for $W=0$. The grey dashed line is the no-cavity reference. \tc{},\td{} Results for an initial cavity excitation with analogous line styles. The parameters are $N=100$, $\lambda=0.4$, and $\nu=0.3g_c$. Averaged over 256 disorder realizations.}
    \label{fig:tail_evolution}
\end{figure}

{\textbf{Reaction coordinate distribution shapes} -- } We are now interested in the time evolution of  the full reaction coordinate distribution $P_n(x_n,t)$ of molecule $n$ and in particular, we will analyze the evolution of its tails in Fig.~\ref{fig:tail_evolution}. In a product state ansatz [Eq.~\eqref{eq:ps_ansatz}], the instantaneous nuclear potential corresponds to a shifted harmonic oscillator. Then, nuclear wave packets of the individual molecules would always stay in a Gaussian shape. Crucially, this is not the case if we allow for finite entanglement. Then, in general, the Holstein coupling $\propto \hat x_n \hat \sigma^+_n \hat \sigma^-_n$ does not factorize and thus modifies the wave packet shape over time (see e.g.~Fig.~\ref{fig:distributions}e). To exemplify this, consider a single molecule $n$ with a constant excitation fraction $\beta$ (without cavity coupling). The time evolution under the Holstein Hamiltonian [Eq.~\eqref{eq:ham-h}] leads to the following state at time $t$: $\ket{\phi_\beta(t)} = \sqrt{1-\beta} \ket{0}_\mathrm{exc}\ket{0}_\mathrm{vib} + \sqrt{\beta} \exp[i\phi(t)] \ket{1}_\mathrm{exc}\ket{x(t)+\mi p (t)}_\mathrm{vib}$ with the coherent state $\ket{\alpha}_\mathrm{vib} = \mathrm{exp} (\alpha \hat b_n^\dag - \alpha^*\hat b_n  )\ket{0}_{\rm vib}$ and the phase $\phi(t)$ due to the energy difference between states $\ket{0}_\mathrm{exc}$ and $\ket{1}_\mathrm{exc}$. For $\beta \neq 0,1$, this is generally an entangled state, and the shape of the nuclear wave packet (after tracing out the spin degree of freedom) is modified from the Gaussian shape, dependent on $\beta$. 
\medskip

In order to numerically study the shape of $P_n(x_n,t)$ with our exact MPS method, we define its tails by $x_n < x_\mathrm{thr}^l$ and $x_n > x_\mathrm{thr}^r$, respectively. Here we choose a threshold value such that the tails of a ground state molecule include one percent of the weight $\eta_0 = \{1 \pm \erf[x_\mathrm{thr}^{l/r}]\}/2 = 10^{-2}$, which corresponds to~$x_\mathrm{thr}^{l/r} \approx \mp 1.6$. We have confirmed that the underlying physics is generally independent of the specific choice of $\eta_0$ for $0.1 > \eta_0 > 10^{-4}$, however the relative magnitude of the changes generally increases for decreasing $\eta_0$. We define the time-dependent tail weights:

\begin{align}
    \eta^{l/r}(t) &= \mp \sum_n \int_{x_\mathrm{thr}^{l/r}}^{\mp\infty} dx_n\,P_n(x_n,t)\,. \label{eq:def_tail}
\end{align}
In a simplified reaction picture, $\eta^{l/r}(t)$ may be related to a reaction probability, e.g.~for dissociation, if the coordinate $x$ corresponds to the stretching of a critical bond in the system~\cite{vendrell2018coherent}.

Without cavity, we can analytically solve the dynamics of the tail weights. For an initial single molecular excitation, the $(N-1)$ ground state molecules exhibit no dynamics, and the excited molecule oscillates between $x_1=0$ and $x_1=2\sqrt{2}\lambda$ according to $x_1(t) = \sqrt{2}\lambda [1 - \cos(\nu t)]$. The tails are then given by $\eta^{l/r}(t) = (N-1) \times \eta_0 + \{1 \pm \erf[x_\mathrm{thr}^{l/r} - x_1(t)]\} / 2$, with $\erf(x) = 2 \int_0^x dz \exp(-z^2) / \sqrt{\pi}$ the error function. This is shown as gray dashed lines in Fig.~\ref{fig:tail_evolution}.

The influence of cavity and disorder on the evolution of $\eta^{l/r}(t)$ is shown in Fig.~\ref{fig:tail_evolution} for both an initial molecule excitation (Fig.~\ref{fig:tail_evolution}a,b) and the cavity excitation scenario (Fig.~\ref{fig:tail_evolution}c,d). We first discuss the disorder-less case $W=0$. For the molecule excitation (light dash-dotted lines in Fig.~\ref{fig:tail_evolution}a/b), we observe only a minimal modification from the no-cavity case (gray dashed line). In contrast, for an initial cavity excitation (Fig.~\ref{fig:tail_evolution}c/d, we find a strong suppression, in particular of the right tail weights due to the cavity. This is a manifestation of the polaron decoupling~\cite{herrera2016cavity}.

For $W=g_c/2$, in contrast, we find a distinctively different behavior. Focusing first on the right tail, we observe that disorder on average leads to a reduction of the tail at $t \sim \pi/\nu$ compared to the no-cavity scenario, followed by an increase at later times, seen in Figs.~\ref{fig:tail_evolution}b,d. This effect is significantly more pronounced for an initial cavity excitation (Fig.~\ref{fig:tail_evolution}d) than for a molecule excitation (Fig.~\ref{fig:tail_evolution}b). We attribute the dynamics observed in Fig.~\ref{fig:tail_evolution}b/d to the out-of-phase oscillation of the different molecular vibrations (cf.~Fig.~\ref{fig:fluctuations}). It implies that reaction coordinates reach large values of $x_n$ at different times and thus reduce the maximum weight of $\eta^r$ at $t\sim\pi/\nu$, but lead to a larger tail weight on average at later times. Importantly, we point again out that this out-of-phase oscillation should be considered as a quantum coherent process, i.e.~the time-dependent state is a large superposition where the vibrational degrees of freedom enter as linear superposition, as for the single molecule state $\ket{\phi_\beta}$, but with molecule and time-dependent excitation fractions. The importance of vibrational entanglement for modeling the exact dynamics of the nuclear distribution is strikingly illustrated by the fact that product state simulations in Fig.~\ref{fig:tail_evolution} (dotted lines) fail to describe the correct dynamics.

We note that when we consider a time integration of the right tail-weights, $\eta^r_\mathrm{avg} = \int_0^{2\pi/\nu} dt\,\eta^r(t)$, i.e.~the surface under the curves in Fig.~\ref{fig:tail_evolution}b/d, we find that for large $W$, the exact ${\eta}^r_\mathrm{avg}$ approximately agrees with the no-cavity scenario. This phenomenon can be rationalized by the fact that, although the excitation is time-dependently distributed over many molecules, in total there still only approximately remains one molecular excitation driving vibrational dynamics. Interestingly, this is not the case when time integrating the left tail weight, ${\eta}^l_\mathrm{avg}$. In fact, this integrated weight increases significantly compared to the no-cavity case, which highlights the importance of the broadening and the non-Gaussian shapes of the nuclear distributions.

\begin{figure}
    \centering
    \includegraphics[width=\columnwidth]{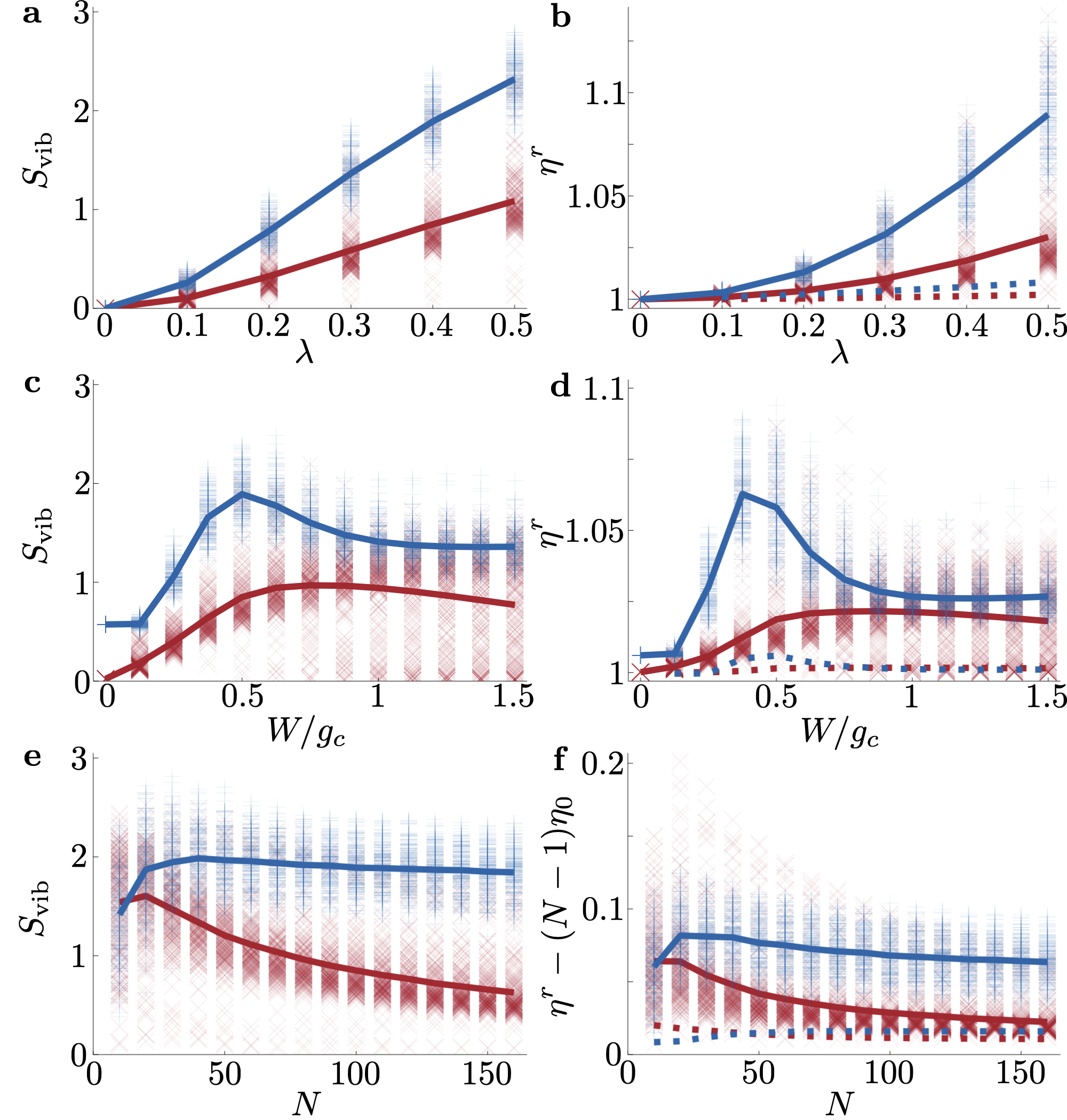}
    \caption{\textbf{Parameter scaling.} The left panels (\ta{}, \tc{}, \te{}) show the entanglement entropy $S_\mathrm{vib}$, while the right panels (\tb{}, \td{}, \tf{}) show the right tail weight $\eta^r$ [defined in Eq.~\eqref{eq:def_tail}] as a function of vibronic coupling strength $\lambda$ (\ta{}, \tb{}), disorder width $W$ (\tc{}, \td{}), and molecule number $N$ (\te{}, \tf{}), evaluated at time $t=2\pi/\nu$. Red ``$\times$'' symbols correspond to a molecular excitation, blue ``$+$'' symbols correspond to a cavity excitation. The symbols represent individual disorder realizations. The continuous line is a guide to the eye through averages of all 256 disorder realizations. The dotted lines in panels b, d, and f represent the disorder averaged product state results for reference. $\eta_0 = 10^{-2}$ is the tail weight in the ground state. Parameters are  $N=100$, $\nu = 0.3g_c$, $\lambda = 0.4$, and $W=g_c/2$ unless specified.}
    \label{fig:wlamn-scaling}
\end{figure}

\medskip

\textbf{Parameter scaling -- } Lastly we want to systematically investigate the importance of the effects introduced in this paper as function of disorder strength $W$, vibronic coupling strength $\propto \lambda$, and molecule number $N$. In Fig.~\ref{fig:wlamn-scaling} we focus on the entanglement entropies and the right tail weights at time $t = 2\pi / \nu$ (red: initial molecule excitation, blue: initial cavity excitation). We find that the entanglement entropy $S_\mathrm{vib}$ and the right tail weight $\eta^r$ scale extremely similarly with all parameters (comparing left a,c,e and right b,d,f panels in Fig.~\ref{fig:wlamn-scaling}, respectively). This confirms the close relation between both quantities. Furthermore, we find that for sufficiently large $\lambda$ and $W$, the product state approximation (dotted lines in Fig.~\ref{fig:wlamn-scaling}) breaks down completely and predicts only negligible modifications compared to the exact MPS simulations. This coincides with large values of $S_\mathrm{vib}$, and thus underlines the essential role of entanglement between electro-photonic and vibrational degrees of freedom in the dynamics.

We observe that both $S_\mathrm{vib}$ and $\eta^r$ grow with $\lambda$ (see Fig.~\ref{fig:wlamn-scaling}a,b. As discussed above, both entanglement and modifications to the right tail can be directly attributed to $\hat H_H$, which scales with $\lambda$ [Eq.~\eqref{eq:ham-h}]. For small disorder $W < g_c/2$, i.e.~in the strong coupling regime, we find that increasing disorder results in an increase of entanglement entropies and right tail weights (Fig.~\ref{fig:wlamn-scaling}c,d), consistent with disorder enhanced excitation transfer. Interestingly, $S_\mathrm{vib}$ and $\eta^r$ exhibit a peak between the weak and strong coupling limits. It becomes only weakly dependent on $W$ in the weak coupling regime, i.e.~for $W>g_c$. This behavior and the clear difference between excitation scenarios exemplifies the rich physics in the intermediate coupling regime.

\medskip

Strikingly, we also observe different scaling behaviors with the molecule number $N$ between both initial states (Fig.~\ref{fig:wlamn-scaling}e,f). For an individually excited molecule, the entanglement and the right tail weight decrease for large $N$ [we subtract the ground state contribution $(N-1)\eta_0$ from the tail weight], in line with the analytical estimate for the scaling of excitation transfer between molecules $\sim Wt/N$ in the strong coupling regime. In contrast, for an initial cavity-excitation, the entanglement and tail weight remain approximately constant for large $N$. Here, the excitation transfer from cavity to molecules occurs on the same timescale as Rabi oscillations, and the total amount of excitation transferred is perturbatively given by $W^2/g_c^2$ in the strong coupling regime, and thus to first order independent of $N$. This further highlights the important distinction between the two initial states, especially for large molecule numbers.

\section{Conclusion and Outlook}

In summary, we have analyzed the coherent femtosecond dynamics in a disordered Holstein-Tavis-Cummings model after incoherent photo-excitation. This minimal model features necessary ingredients for analyzing key quantum processes in polaritonic chemistry, including dynamics of electronic, vibrational, and photonic degrees of freedom~\cite{herrera2016cavity}. Using a matrix product state approach we have simulated the exact quantum many-body dynamics for realistic parameter regimes for mesoscopic system sizes. We have shown that disorder-enhanced excitation transfer~\cite{botzung2020dark,chavez2021disorder,Dubail2021}, both between the molecules and from the cavity to molecules, leads to coherent out-of-phase oscillations of the individual vibrational modes. Disorder thus strongly enhances the build-up of vibrational entanglement and modifications of the time-dependent nuclear probability distributions, which are not captured in a product state (mean-field Ehrenfest) picture where electronic and nuclear degrees of freedom are treated as separable. We have highlighted that for large molecule numbers, an initial excitation in the cavity leads to much larger modifications than an initial molecular excitation. In general, disorder-enhanced entanglement is a remarkable effect, since typically disorder is known to lead to a suppression of entanglement in various quantum many-body models~\cite{Abanin_Collo_2019}.

Our results have direct implications for understanding the role of collective and quantum-mechanical effects in cavity modified chemistry. While approximations based on wave-functions which are separable between the electronic and the vibrational Hilbert space can provide useful insight in disorder-free systems, our work implies that the presence of disorder leads to a breakdown of such approaches. Our work emphasizes that for cavity-modified photo-chemistry with incoherent excitations, it is crucial to distinguish scenarios where the cavity or individual molecules are activated by the photon. The observation of large-scale entanglement entropy build-up on very short femtosecond timescales suggests that quantum effects can play an important role in polaritonic chemistry experiments on timescales faster then the cavity-decay, and thus for experimentally feasible cavities with quality factors of $Q\gtrsim 1000$. Our work highlights the general importance of disorder for understanding polaritonic chemistry~\cite{scholes2020polaritons,sommer2020molecular,du2021can}.

In the future it will be interesting to consider more realistic molecular models, including beyond-harmonic potential energy landscapes with more than one reaction coordinate, and featuring chemical reactions e.g.~via electron transfer between multiple electronic levels, or conical intersections of energy surfaces. It will be interesting to extend our analysis to much longer times, when disorder enhanced transfer becomes even more relevant~\cite{thomas_these}. Our numerical approach further allows to also access regimes with multiple excitations, which will be an interesting regime to explore. Furthermore our method can also easily include dissipative mechanisms, e.g.~using a quantum trajectory approach~\cite{Wall_Simul_2016}, which has been proposed to lead to further modifications of the involved chemistry~\cite{Felicetti2020Oct,Antoniou2020Oct,Wellnitz2021Feb,Torres-Sanchez2021Jan,Davidsson2020Dec}, an interesting prospect for future research.

{\textbf{Acknowledgements} --- }
We are grateful to Felipe Herrera, Claudiu Genes, David Hagenm\"uller, Jer\^ome Dubail and Guido Masella for stimulating discussions. This work was supported by LabEx NIE (``Nanostructures in Interaction with their Environment'') under contract ANR-11-LABX0058 NIE and ``ERA-NET QuantERA'' - Projet ``RouTe'' (ANR-18-QUAN-0005-01). This work of the Interdisciplinary Thematic Institute QMat, as part of the ITI 2021 2028 program of the
University of Strasbourg, CNRS and Inserm, was supported by IdEx Unistra (ANR 10 IDEX 0002), SFRI STRAT’US project (ANR 20 SFRI 0012), and EUR QMAT ANR-17-EURE-0024 under the framework of the French
Investments for the Future Program. G. P. acknowledges support from the Institut Universitaire de France (IUF) and the University of Strasbourg Institute of Advanced Studies (USIAS). Our MPS codes make use of the intelligent tensor library (ITensor)~\cite{itensor}. Computations  were  carried  out  using  resources  of  the High Performance Computing Center of the University of Strasbourg, funded by Equip@Meso (as part of the Investments for the Future Program) and CPER Alsacalcul/Big Data.

\bibliography{main}

\clearpage
\newpage

\begin{widetext}

\section*{Appendix}

\subsection{Matrix product state method}

We write the time-dependent quantum state on the full electro-photonic-vibrational Hilbert space in the form

\begin{align}
    \ket{\psi(t)} = &\sum_{\{i_n=0,1\}} \sum_{\{b_n=0\}}^{n_{\rm max}^v} \sum_{a=0}^{n_{\rm max}^p} c_{i_1,i_2,\dots,i_N; b_1,b_2,\dots,b_N; a}\\
    &\quad \ket{i_1,i_2,\dots,i_N}_{\rm exc} \otimes \ket{b_1,b_2,\dots,b_N}_{\rm vib} \otimes \ket{a}_{\rm ph} \nonumber
\end{align}
Here, the different indices denote electronic excitation numbers for molecule $n$, $i_n=0,1$, the vibrational excitation number on molecule $n$, $b_n=0,\dots, n_{\rm max}^v$ and the cavity mode occupation number, $a=0,\dots,n_{\rm max}^p$. While in principle $n_{\rm max}^{v/p} \to \infty$, in practice the vibrational Hilbert space can be truncated at some reasonable occupation number. For this work we found that to capture all relevant physics of the tails of the nuclear coordinate distributions, $n_{\rm max}^v = 10$ is sufficient~(see below). Due to our choice of initial state and the conservation of $\sum_n \hat \sigma_n^+ \hat \sigma_n^- + \hat a^\dag \hat a $, furthermore we can set a photon cutoff at $n_{\rm max}^v = 1$ without any approximation. For $N=100$ molecules, this implies a full Hilbert space size of $11^N 2^{N+1} \gtrsim 10^{134}$, clearly out of reach for any classical computer memory. In order to still make the high-dimensional complex state tensor $c$ amenable for storage in computer memory, we utilize a decomposition into products of smaller tensors, a matrix product state (MPS)~\cite{Schollwock_Density-matrix_2011}. In particular we utilize an MPS with $2N+1$ tensors:

\begin{align}
    &c_{i_1,i_2,\dots,i_N; b_1,b_2,\dots,b_N; a} = \\
    &\quad \sum_{\alpha_1,\dots, \alpha_{2N}=1}^{\chi} \Gamma_a^{[p]; \alpha_0,\alpha_1} \prod_{n=1}^N\left(\Gamma_{i_n}^{[n]; \alpha_{2n-1}\alpha_{2n}} \Gamma_{b_n}^{[v]; \alpha_{2n}\alpha_{2n+1}}\right). \nonumber
\end{align}
Here we introduced 3-dimensional tensors for the photonic, electronic and vibrational degrees of freedom, $\Gamma_a^{[p]; \alpha_0\alpha_1}$, $\Gamma_{i_n}^{[n]; \alpha_m\alpha_{m+1}}$, and $\Gamma_{b_n}^{[v]; \alpha_{m}\alpha_{m+1}}$, respectively. The tensors are connected by the virtual indices $\alpha_m$ with $m=0,\dots,2N+1$ and bond dimension $\chi$ (except for the edge indices, which are trivially $\alpha_0=\alpha_{2N+1}=1$). The MPS can be brought, and updated, in a canonical form. Then, the virtual indices $\alpha_m$ correspond to an orthonormal basis, which is the eigenbasis of the reduced density matrix of the two blocks that the index connects~\cite{Schollwock_Density-matrix_2011}. This effectively limits the entanglement entropy between the two blocks to $< \log_2(\chi)$. For the MPS decomposition to become exact, one would need to choose very large values for $\chi \sim \exp(N)$. However, limiting $\chi$ to computationally treatable magnitudes allows to effectively simulate dynamics on a truncated Hilbert space with restricted entanglement. In our simulations, we verified that all results converge with increasing $\chi$, and therefore that our simulations capture all necessary entanglement and are quasi-exact. In practice, we use $\chi=128$ for all plots (see below for convergence plots). In our MPS form, tensors can be updated using the time-evolving Block decimation (TEBD) algorithm~\cite{Vidal_Effic_2004}. Then, HTC coupling terms can be incorporated with ``nearest-neighbor'' gate updates, while cavity-couplings can be incorporated using index-swap gates between the tensors and nearest-neighbor gates. In practice, we choose a second order TEBD decomposition of the Hamiltonian with a time step of $g_c/100$, which we have verified to be sufficiently small for errors due to a finite time step to be negligible (see \appname{}). In case of spin-boson dynamics, TEBD in combination with swap gates have been previously shown to exhibit very well behaved convergence, which are preferable compared to updates that use variational concepts~\cite{Wall_Simul_2016}. Similarly, in order to compute $S_{\rm vib}$ we re-organize all vibrational degrees of freedom into a single block (using swap gates) and compute the entropy over the virtual index into that block. The excitation number conservation can be exploited to enhance the efficiency of tensor contractions and decompositions.

\subsection{Dark state contribution to $\ket{\psi_0^m}$ without disorder}
The eigenstates of $\hat H_\mathrm{TC}$ are given by two polaritons $\ket{\pm} = \hat a^\dagger / \sqrt{2} \pm \sum_n \hat \sigma_n^+ / \sqrt{2N}\ket{0}_\mathrm{exc+ph}$, and $N-1$ degenerate dark states, for which any orthonormal basis of $\mathcal H_\mathrm{ph} \otimes \mathcal H_\mathrm{exc}$ that does not contain the polaritons may be chosen. The most straightforward way to compute the dark state contribution of a state $\ket \psi$ is thus $\sum_d \abs{\bra{d}\ket{\psi}}^2 = 1 - \abs{\bra{+}\ket{\psi}}^2 - \abs{\bra{-}\ket{\psi}}^2$. We find $\sum_d \lvert\bra{d}\ket{\psi_0^m}\rvert^2 = 1 - 1/N$.

\subsection{Perturbative photo-contribution to the dark states}

In this section, we compute the photo-contribution to the dark states perturbatively in the regime $g_c \gg W, \nu, \lambda\nu$. Starting from the analytically solvable Hamiltonian $\hat H_0 = \hat H_\mathrm{TC} + \hat H_\mathrm{vib}$ given by Eqs.~\eqref{eq:ham-tc} and \eqref{eq:ham-vib} of the main text, we compute the perturbative corrections to the photo-contribution of the polaritons. The perturbative photo-contribution to the dark states can then be computed as $\sum_d \lvert\bra{d}\ket{1}_\mathrm{ph}\rvert^2 = 1 - \lvert\bra{+}\ket{1}_\mathrm{ph}\rvert^2 - \lvert\bra{-}\ket{1}_\mathrm{ph}\rvert^2$, analogous to the previous section.

The eigenstates of $\hat H_0$ are product states of the eigenstates of $\hat H_\mathrm{TC}$ and the eigenstates of $\hat H_\mathrm{vib}$, because the two sub-spaces remain uncoupled. The eigenstates of $\hat H_\mathrm{TC}$ are the two polaritons $\ket{\pm}$ and the $N-1$ degenerate dark states, for which we choose a momentum basis $\ket{k} = \big[\sum_{n=1}^N \exp(-2\pi\mi kn/N) \hat \sigma_n^+ / \sqrt{N}\big] \ket{0}_\mathrm{exc+ph}$. The eigenstates of $\hat H_\mathrm{vib}$ are Fock states $(\prod_n \hat (b_n^\dagger)^{i_n} / \sqrt{i_n!}) \ket{0}_\mathrm{vib}$ with $i_n$ vibrational excitations on the $n$-th molecule.

\subsubsection{Disorder}
We first compute the perturbative corrections due to disorder $\hat H_\mathrm{dis}$, only. In this case, the vibrations are not entangled with the photo-electronic degrees of freedom, and we can restrict the perturbative analysis to the photo-electronic degrees of freedom, only. The second order corrections to the states $\ket{\pm}$ are given by~\cite{sakurai1995modern}
\begin{align}
    \ket{\pm} &= \ket{\pm^{(0)}} + \ket{\pm^{(1)}} + \ket{\pm^{(2)}} + \mathcal O(W^3/g_c^3) \label{eq:perturbation-theory0} \\
    \bra{\psi^{(0)}} \ket{\phi^{(1)}} &=
    \left\{
        \begin{matrix}
            0 &\quad \text{for $\psi = \phi$} \\
            \frac{\bra{\psi^{(0)}} \hat H_\mathrm{dis} \ket{\phi^{(0)}}}{E_\phi^{(0)} - E_\psi^{(0)}} &\quad \text{else}
        \end{matrix}
    \right.\\
    \bra{\psi^{(0)}} \ket{\phi^{(2)}} &= 
    \left\{
        \begin{matrix}
            - \frac{1}{2} \sum_{\eta \neq \phi} \frac{\bra{\psi^{(0)}} \hat H_\mathrm{dis} \ket{\eta^{(0)}} \bra{\eta^{(0)}} \hat H_\mathrm{dis} \ket{\phi^{(0)}} }{\qty(E_\phi^{(0)} - E_\eta^{(0)})^2}& \quad \text{for $\psi = \phi$} \\
            \sum_{\eta \neq \phi} \frac{\bra{\psi^{(0)}} \hat H_\mathrm{dis} \ket{\eta^{(0)}} \bra{\eta^{(0)}} \hat H_\mathrm{dis} \ket{\phi^{(0)}} }{\qty(E_\phi^{(0)} - E_\eta^{(0)}) \qty(E_\phi^{(0)} - E_\psi^{(0)})}
            - \frac{\bra{\psi^{(0)}} \hat H_\mathrm{dis} \ket{\phi^{(0)}} \bra{\phi^{(0)}} \hat H_\mathrm{dis} \ket{\phi^{(0)}} }{\qty(E_\phi^{(0)} - E_\psi^{(0)})^2} & \quad \text{else}
        \end{matrix}
    \right. \label{eq:perturbation-theory1}
\end{align}
with $\ket{\psi^{(0)}}$ and $E_\psi^{(0)}$ the eigenstates and eigenenergies for $W=0$, and $\ket{\psi^{(n)}}$ the corrections to the eigenstates at order $(W / g_c)^n$.
We find
\begin{align}
    \bra{\mp^{(0)}}\ket{\pm^{(1)}} &= \mp\frac{\tilde{\epsilon}_0}{4g_c} \, , \\
    \bra{k^{(0)}}\ket{\pm^{(1)}} &= \pm \frac{\tilde{\epsilon}_k}{\sqrt{2}g_c} \, , \\
    \bra{\pm^{(0)}}\ket{\pm^{(2)}} &= - \frac{\abs{\tilde{\epsilon}_0}^2}{32g_c^2} - \sum_{k=1}^{N-1} \frac{\abs{\tilde{\epsilon}_k}^2}{4g_c^2} \, , \\
    \bra{\mp^{(0)}}\ket{\pm^{(2)}} &= - \sum_{k=1}^{N-1} \frac{\abs{\tilde{\epsilon}_k}^2}{4g_c^2} \, ,
\end{align}
with $\tilde{\epsilon}_k = [\sum_{n=1}^N \exp(2\pi \mi k n/N) \epsilon_n]/N$ the Fourier transform of the random energies. Using $\ket{1_\mathrm{ph}} = (\ket{+^{(0)}}+\ket{-^{(0)}})/\sqrt{2}$, we find further
\begin{align}
    \abs{\bra{1_\mathrm{ph}}\ket{\pm}}^2 = \frac{1}{2} \mp \frac{\tilde{\epsilon}_0}{4g_c} - \frac{1}{2} \sum_{k=1}^{N-1} \frac{\abs{\tilde{\epsilon}_k}^2}{g_c^2} + \mathcal{O}\qty[(\tilde \epsilon/g_c)^3] \, .
\end{align}
and finally for the photon weight of the dark states
\begin{align}
    \sum_d \abs{\bra{1_\mathrm{ph}}\ket{d}}^2 = \sum_{k=1}^{N-1} \frac{\abs{\tilde \epsilon_k}^2}{g_c^2} \, .
\end{align}
Until here, the results are independent of the specific disorder model. We now assume that the energies $\epsilon_i$ are distributed according to a Gaussian with mean zero and standard deviation $W$. Then, we can further use that the discrete Fourier transform of a set of $N$ Gaussian random variables is a set of $N$ complex Gaussian random variables with real and imaginary part mean zero and standard deviation $W/\sqrt{2N}$, as can be straightforwardly shown by Fourier transforming the definition of $P(\tilde \epsilon_k)$. Taking the disorder average leaves us with
\begin{align}
    \overline{\sum_d \abs{\bra{1_\mathrm{ph}}\ket{d}}^2} = \frac{(N-1)W^2}{Ng_c^2} + \mathcal{O}(W^3/g_c^3)  \xrightarrow{N \rightarrow \infty} \frac{W^2}{g_c^2} \, .
\end{align}

\subsubsection{Vibronic coupling}
We now proceed to compute the corrections due to vibronic coupling $\hat H_\mathrm{H}$ for $W=0$. $\hat H_\mathrm{H}$ couples states with different numbers of vibrations, so that we need to take the vibrations into account explicitly. We write the states as $\ket\psi = \ket{\psi_\mathrm{exc+ph}, n_\mathrm{vib}, v_1, \dots, v_{n_\mathrm{vib}}} \equiv \mathcal{N} \ket{\psi_\mathrm{exc+ph}} \prod_{i=1}^{n_\mathrm{vib}} \sum_n \exp(2\pi\mi v_in/N)\hat b_n^\dagger \ket{0}_\mathrm{vib}$, where $\psi_\mathrm{exc+ph} = \pm, k$ is the state in the electro-photonic sub-space, $n_\mathrm{vib}$ is the number of vibrations, and $v_i$ is the Fourier mode of the vibration and $\mathcal N$ is a normalization factor. We are specifically interested in the contribution of the initial state $\ket{\psi_0^c} = \ket{1_\mathrm{ph}, 0}$ to the dark states. Plugging  $\hat H_\mathrm{H}$ instead of $\hat H_\mathrm{dis}$ into Eqs.~\eqref{eq:perturbation-theory0} to \eqref{eq:perturbation-theory1}, we find
\begin{align}
    \bra{\pm^{(0)},1,0}\ket{\pm^{(1)}, 0} &= \frac{\lambda}{2\sqrt{N}} \, , \\
    \bra{\mp^{(0)},1,0}\ket{\pm^{(1)}, 0} &= \frac{\pm\lambda\nu}{2\sqrt{N}\qty(2g_c \mp \nu)} \, , \\
    \bra{k^{(0)},1,-k}\ket{\pm^{(1)}, 0} &= \frac{-\lambda\nu}{\sqrt{2N}\qty(g_c \mp \nu)} \, , \\
    \bra{\pm^{(0)}, 0}\ket{\pm^{(2)}, 0} &= - \frac{\lambda^2}{8N} - \frac{\lambda^2\nu^2}{8N\qty(2g_c \mp \nu)^2} - \frac{(N-1)\lambda^2\nu^2}{4N\qty(g_c \mp \nu)^2} \, , \\
    \bra{\mp^{(0)}, 0}\ket{\pm^{(2)}, 0} &= \pm \frac{\lambda^2\nu}{8Ng_c} - \frac{\lambda^2\nu^2}{8Ng_c(2g_c \mp \nu)} - \frac{(N-1)\lambda^2\nu^2}{4N\qty(g_c \mp \nu)g_c} \, , \\
    \bra{\pm^{(0)}, 0} \ket{\pm^{(2)}, 2, 0, 0} &= \frac{\sqrt{2}\lambda^2}{8N} \pm \frac{\sqrt{2}\lambda^2\nu}{8N(2g_c \pm \nu)} \, , \\
    \bra{\pm^{(0)}, 0} \ket{\pm^{(2)}, 2, k, -k} &= \pm \frac{\lambda^2\nu}{4N(g_c \pm \nu)} \, , \\
    \bra{\mp^{(0)}, 0} \ket{\pm^{(2)}, 2, 0, 0} &= \mp \frac{\sqrt{2}\lambda^2\nu}{8N(g_c \pm \nu)} - \frac{\sqrt{2}\lambda^2\nu^2}{8N(2g_c \pm \nu)(g_c \pm \nu)} \, , \\
    \bra{\mp^{(0)}, 0} \ket{\pm^{(2)}, 2, k, -k} &= -\frac{\lambda^2\nu^2}{4N(g_c \pm \nu)^2} \, , \\
    \bra{\pm^{(0)}, 0} \ket{\pm^{(1)}, 1, 0} &= -\frac{\lambda}{2\sqrt{N}} \, , \\
    \bra{\mp^{(0)}, 0} \ket{\pm^{(1)}, 1, 0} &= \pm \frac{\lambda\nu}{2\sqrt{N}\qty(2g_c \pm \nu)} \, .
\end{align}
Analogous to above, we can compute the dark state amplitude of the state $\ket{1_\mathrm{ph},0}$ including an arbitrary number of vibrations as
\begin{align}
    \sum_d \abs{\bra{1_\mathrm{ph},0}\ket{d}}^2 
    = 1 
    &- \abs{\bra{1_\mathrm{ph},0}\ket{+,0}}^2 
    - \abs{\bra{1_\mathrm{ph},0}\ket{+,1,0}}^2 
    - \abs{\bra{1_\mathrm{ph},0}\ket{+,2,0,0}}^2 
    - \sum_k\abs{\bra{1_\mathrm{ph},0}\ket{+,2,k,-k}}^2 \nonumber \\
    &- \abs{\bra{1_\mathrm{ph},0}\ket{-,0}}^2 
    - \abs{\bra{1_\mathrm{ph},0}\ket{-,1,0}}^2 
    - \abs{\bra{1_\mathrm{ph},0}\ket{-,2,0,0}}^2 
    - \sum_k\abs{\bra{1_\mathrm{ph},0}\ket{-,2,k,-k}}^2 \, .
\end{align}
Using $\ket{1_\mathrm{ph},0} = ( \ket{+^{(0)}, 0} + \ket{-^{(0)}, 0} ) / \sqrt{2}$, the only non-vanishing term for large $N$ reads
\begin{align}
    \abs{\bra{1_\mathrm{ph}, 0}\ket{\pm, 0}}^2 &=  
    \frac{1}{2} 
    - \frac{\lambda^2\nu^2}{4\qty(g_c \mp \nu)^2} 
    - \frac{\lambda^2\nu^2}{4\qty(g_c \mp \nu)g_c} + \mathcal{O}\qty[(\lambda\nu/g_c)^3] \, .
\end{align}
As a result, the photo-contribution to the dark states is approximately
\begin{align}
    \sum_d \abs{\bra{1_\mathrm{ph}}\ket{d}}^2 \approx
    \frac{\lambda^2\nu^2}{4\qty(g_c \mp \nu)^2} 
    + \frac{\lambda^2\nu^2}{4\qty(g_c \mp \nu)g_c} \approx \frac{\lambda^2\nu^2}{2g_c^2} \, ,
\end{align}
where we used $\nu \ll g_c$ in the last step.

\subsection{Excitation Transfer Estimates}
In the following, we derive analytical estimates for excitation transfer away from the initially excited molecule for the initial state $\ket{\psi_0^m} = \hat \sigma_1^+ \ket{0}$ (see main text). We consider disorder induced transfer only, i.e.~we set $\lambda = 0$. We further restrict the analysis to the perturbative case by assuming $W \ll g_c$.

\subsubsection{$W=0$}
We start by analyzing the disorder-free scenario. For $W=\lambda=0$, the electro-photonic state evolves with the Tavis-Cummings Hamiltonian $\hat H_\mathrm{TC}$, only. By diagonalizing $\hat H_\mathrm{TC} = (g_c \ket{+}\bra{+} - g_c \ket{-}\bra{-})$, we compute the time-dependent excitation probability of the initially excited molecule
\begin{align}
    \langle \hat \sigma_1^+ \hat \sigma_1^- \rangle(t) &= \bra{0} \hat\sigma_1^- \exp( \mi \hat H_\mathrm{TC} t) \hat \sigma_1^+ \hat \sigma_1^- \exp( - \mi \hat H_\mathrm{TC} t) \hat \sigma_1^+ \ket{0} \nonumber \\
    &= \frac{(N-1)^2}{N^2} + \frac{2(N-1)}{N^2} \cos(g_ct) + \frac{1}{2N^2} + \frac{1}{2N^2} \cos(2g_ct) \\
    &= 1 + \frac{2}{N} \qty[\cos(g_ct) - 1] + \mathcal{O}\qty(\frac{1}{N^2})\, .
\end{align}

\subsubsection{$W>0$}

In the following we compute the excitation probability of the initially excited molecule for finite disorder. Here, we choose a box disorder model, because the finite probability for a single molecule to have extreme energies (i.e.~$\epsilon_i > +g_c$ or $\epsilon_i < -g_c$) adds significant complexity to the analytical treatment. In particular, we assume a uniform probability $P(\epsilon_i) = 1/(2W)$ for $-W < \epsilon_i < W$, and $P(\epsilon_i) = 0$ otherwise. We do not expect this choice for $P(\epsilon_i)$ to modify the overall scaling of our results with respect to the Gaussian choice in the rest of the paper.

In order to compute the time-dependent excitation probability of the first molecule, we treat the cavity coupling of the first molecule $g (\hat \sigma_1^+ \hat a + \hat \sigma_1^- \hat a^\dagger)$ as a perturbation. The unperturbed Hamiltonian is given by $\hat H_0 = \hat H_\mathrm{TC} + \hat H_\mathrm{dis} - g (\hat \sigma_1^+ \hat a + \hat \sigma_1^- \hat a^\dagger)$. In this case, the perturbation condition becomes that the single molecule coupling $g$ is smaller than the energy differences between different states, which is on the order of $\sim W$ for most pairs of states. As $g = g_c/\sqrt{N}$ becomes very small for large $N$, this condition is fulfilled for most energy levels. However, a few resonant energy levels do typically not fulfill this condition, with implications discussed below.

One eigenstate of $\hat H_0$ is the initial state $\ket{1^{(0)}} \equiv \ket{\psi_0^m}$. For $W < g_c$, the other eigenstates can be classified as polaritons and dark states. Although we cannot compute these states exactly, we can make sufficient statements about their statistics~\cite{Dubail2021,botzung2020dark,thomas_these} to compute the time evolution of $\langle \hat \sigma_1^+ \hat \sigma_1^- \rangle$. In particular, we can perturbatively compute their average photo contribution for $W < g_c$ and large $N$ as in the previous section. We find $\abs{\bra{1_\mathrm{ph}}\ket{+}} \approx 1/2 - W^2/(24g_c^2)$ and $\abs{\bra{1_\mathrm{ph}} \ket{d_i} }^2 \sim W^2/(12Ng_c^2)$, where the additional factor of $12$ comes in due to the different disorder model. The eigenstates of $\hat H_0$ are thus
\begin{align}
    \ket{1^{(0)}} &= \hat \sigma_1^+ \ket{0} \, ,\\
    \ket{\pm^{(0)}} &= \qty[\sqrt{\frac{1-W^2/(12g_c^2)}{2}} \hat a^\dagger \pm \sum_{n=2}^N b^\pm_n \hat \sigma_n^+] \ket{0} \, ,\\
    \ket{d_i^{(0)}} &= \qty[\frac{W}{\sqrt{12}Ng}\hat a^\dagger + \sum_{n=2}^N c^{(i)}_n \hat \sigma_n^+] \ket{0} \, ,
\end{align}
where we took the disorder-average of the photon-contribution on the state level. This approximation may be valid due to self-averaging for sufficiently large $N$, and it is validated by the agreement of the final results with the numerical simulations. The $b_n^\pm$ and $c_n^{(i)}$ are constants that determine the excitation probability of the specific molecules and are not needed in the following. The corresponding eigenenergies are
\begin{align}
    E_1^{(0)} &= \epsilon_1 \, , \\
    E_\pm^{(0)} &= \pm g_c \, , \\
    E_{d,i}^{(0)} &\equiv E_i \, ,
\end{align}
where the dark state energies $E_i$ follow the same distribution as the random molecular excitation energies~\cite{Dubail2021}.

Importantly, all states $\ket{\pm^{(0)}}$ and $\ket{d_i^{(0)}}$ have no excitation probability for the first molecule. As a result, there are no corrections to the eigenenergies at first order. The perturbative corrections to the states are
\begin{align}
    \ket{1^{(1)}} &= \frac{\sqrt{1-W^2/(12g_c^2)}g}{\sqrt{2}\qty(\epsilon_1 - g_c)} \ket{+^{(0)}} +
    \frac{\sqrt{1-W^2/(12g_c^2)}g}{\sqrt{2}\qty(\epsilon_1 + g_c)} \ket{-^{(0)}}
    + \sum_i \frac{W}{\sqrt{12}N \qty(\epsilon_1 - E_i)} \ket{d_i^{(0)}} \,, \\
    \ket{d_i^{(1)}} &= \frac{W}{\sqrt{12}N\qty(E_i - \epsilon_1)} \ket{1^{(0)}} \,, \\
    \ket{\pm^{(1)}} &= \frac{\sqrt{1-W^2/(12g_c^2)}g}{\sqrt{2}\qty(\pm g_c - \epsilon_1)} \ket{1^{(0)}} \,, \\
    \bra{1^{(0)}}\ket{1^{(2)}} &= - \frac{1}{2} \bra{1^{(1)}}\ket{1^{(1)}} \, .
\end{align}
Here, unphysical divergences appear for the resonance condition $\epsilon_1 \rightarrow E_i$. These are related to the perturbation assumption $g \ll \abs{\epsilon_1 - E_i}$. We will deal with these divergences below.

The time evolution is computed as
\begin{align}
    \langle \hat \sigma_1^+ \hat \sigma_1^- \rangle (t) &= \bra{\psi_0^m} \exp(\mi \hat H t) \hat \sigma_1^+ \hat \sigma_1^- \exp(-\mi \hat H t) \ket{\psi_0^m} \nonumber \\
    &= \bra{1^{(0)}} \exp(\mi \hat H t) \ket{1^{(0)}} \bra{1^{(0)}} \exp(-\mi \hat H t) \ket{1^{(0)}} \, .
\end{align}
We expand $\mathrm{exp}(\mi \hat H t) = \sum_\psi \ket{\psi}\bra{\psi} \exp(\mi E_\psi t)$ and keep terms only up to second order in $W$. We furthermore ignore second order corrections to the energies, which would lead to higher order corrections of the final result. We find
\begin{align}
    \bra{1^{(0)}} \ket{1} \exp(\mi\epsilon_1t) \bra{1} \ket{1^{(0)}} &= 
        \qty[1 - 
        \frac{g_c^2 - W^2/12}{2N\qty(\epsilon_1 - g_c)^2} - 
        \frac{g_c^2 - W^2/12}{2N\qty(\epsilon_1 + g_c)^2} - 
        \sum_i \frac{W^2}{12N^2 \qty(\epsilon_1-E_i)^2}] 
        \exp(\mi\epsilon_1t) \, ,\\
    \bra{1^{(0)}} \ket{\pm} \exp(\pm\mi g_ct) \bra{\pm} \ket{1^{(0)}} &= 
        \frac{g_c^2 - W^2/12}{2N(\pm g_c - \epsilon_1)^2}
        \exp(\pm\mi g_ct)  \, , \\
    \bra{1^{(0)}} \ket{d_i} \exp(\mi E_i t) \bra{d_i} \ket{1^{(0)}} &=
        \frac{W^2}{12N^2 \qty(\epsilon_1-E_i)^2}
        \exp(\mi E_i t) \, .
\end{align}
These terms lead to oscillations of the energy between the initially excited molecule and the other states at frequencies $g_c \pm \epsilon_1$ for the polaritons, and $E_i - \epsilon_1$ for the dark states, respectively. The combined effect of the slightly out-of-phase oscillations of the large number of dark states leads to an effective dephasing and a resulting unidirectional transfer of energy away from the initially excited molecule on timescales analyzed in the paper. This behavior can be computed as
\begin{align}
    \sum_i \bra{1^{(0)}} \ket{1} \exp(\mi\epsilon_1t) \bra{1} \ket{1^{(0)}} \bra{1^{(0)}} \ket{d_i} \exp(\mi E_i t) \bra{d_i} \ket{1^{(0)}} + \text{h.c.} &=
        \sum_i \frac{W^2}{6N^2 \qty(E_i-\epsilon_1)^2} \cos[\qty(E_i-\epsilon_1)t] \, .
\end{align}
We finally take the disorder average $E_i\rightarrow \int dE / W$ to find:
\begin{align}
    N\times \frac{W}{6N^2}\int_{-W/2}^{W/2} dE \frac{\cos[(E-\epsilon_1)t]}{(E-\epsilon_1)^2} \quad = 
    \quad \frac{Wt}{6N}\int_{(-W/2-\epsilon_1)t}^{(W/2-\epsilon_1)t} d\Delta \frac{\cos(\Delta)}{\Delta^2} \, ,
\end{align}
where we substituted $(E-\epsilon_1)t \rightarrow \Delta$. Note that again an unphysical divergence arises due to the perturbation assumption $g^2 \ll (E_i - \epsilon_1)^2$, which we ignore here as we are only interested in the overall scaling behavior. As the integrand scales like $1/\Delta^2$, the integral becomes time independent for diverging boundaries, i.e.~sufficiently large $t$. In this case, the population of the first state evolves like $1-\langle \hat \sigma_1^+\hat\sigma_1^- \rangle \sim Wt/N$ in addition to Rabi-oscillations.
By straightforwardly evaluating all other terms we find that this term is indeed the largest contribution to the energy transfer. Botzung \textit{et al.}~\cite{botzung2020dark} derived similar results in the long time limit for $N\rightarrow \infty$.

\subsection{Convergence of MPS simulations}

\begin{figure}
    \centering
    \includegraphics[width=0.5\textwidth]{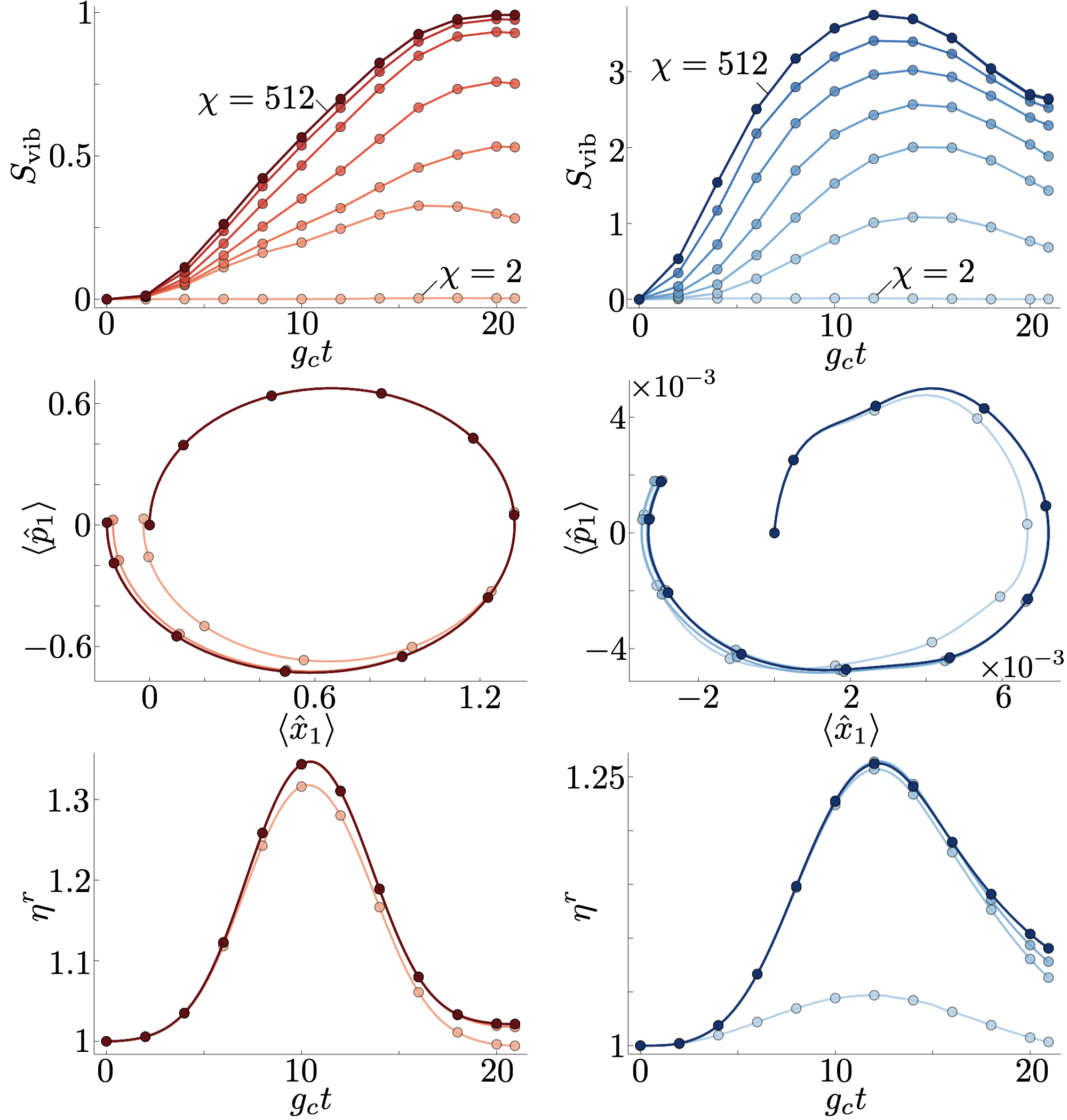}
    \caption{\textbf{Convergence with the bond dimension $\chi$.} \ta{},\tb{} Time evolution of the vibrational entropy $S_\mathrm{vib}$ for an initial molecular (left, red) or cavity (right, blue) excitation. Darkness indicates $\chi$ on a log-scale ($\chi \in \{2,4,8,16,32,64,128,256,512\}$). Trajectories already fully overlap for for $\chi \geq 128$ indicating convergence. In \ta{},\tb{}, the lines are a guide to the eye, whereas in \tc{}-\tf{} the lines represent additional data points. \tc{},\td{} Phase-space evolution of a single molecule. Same color-code as in \ta{},\tb{}. \te{},\tf{} Time evolution of the right tail weight $\eta^r$ (see paper for definition). Same color-code as in \ta{},\tb{}. Parameters for all plots: $N=100$, $\lambda=0.5$, $\nu=0.3g_c$, $W=g_c/2$, $dt=0.01$, $n_\mathrm{max}^v = 10$. Results for a single disorder realization.}
    \label{fig:chiconv}
\end{figure}

\begin{figure}
    \centering
    \includegraphics[width=0.5\textwidth]{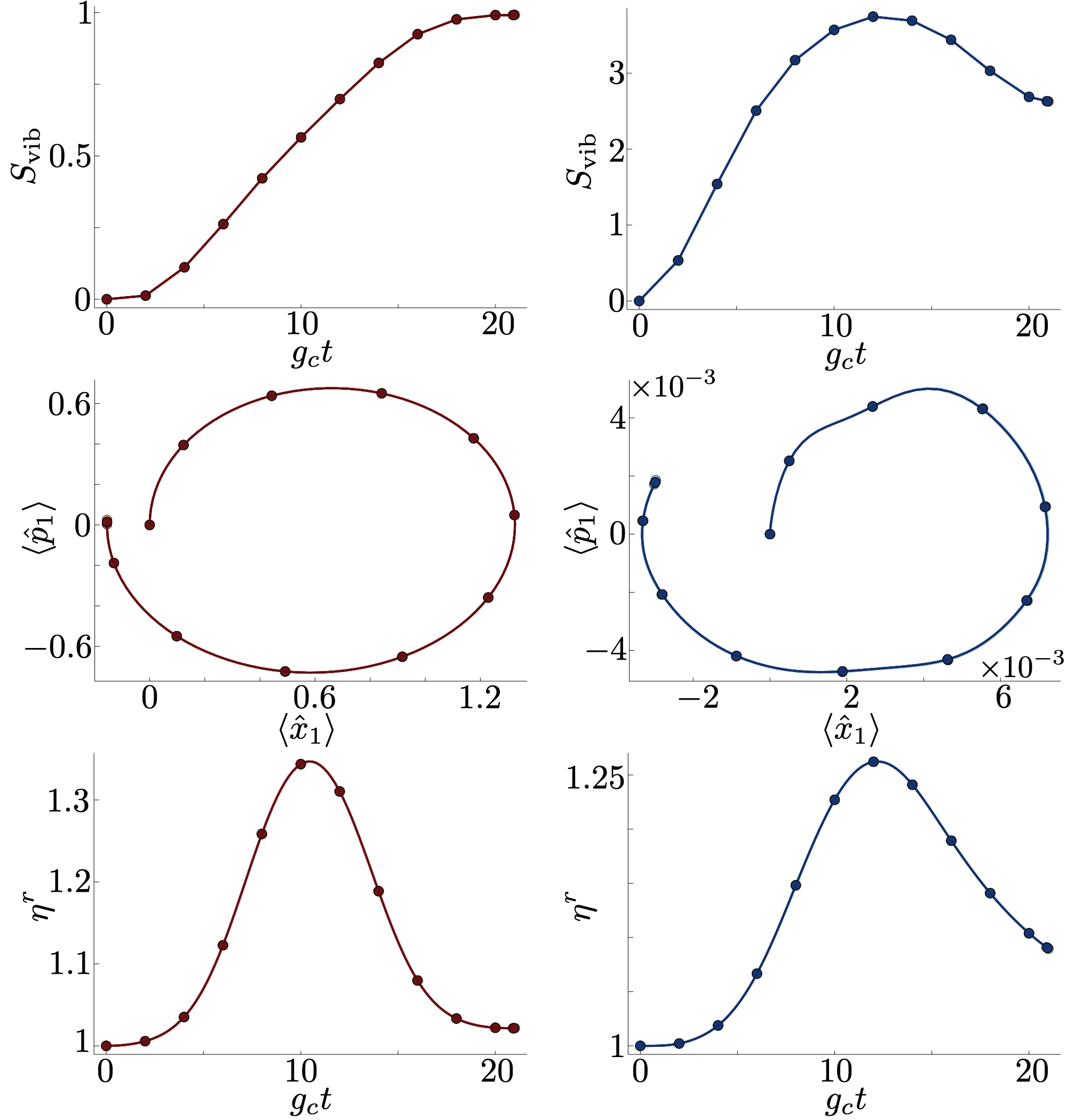}
    \caption{\textbf{Convergence with the time step $dt$ of the simulation.} \ta{},\tb{} Time evolution of the vibrational entropy $S_\mathrm{vib}$ for an initial molecular (left, red) or cavity (right, blue) excitation. Darkness indicates $dt$ on a log-scale ($dt \times g_c \in \{0.2,0.1,0.04,0.02,0.01,0.005,0.0025\}$). Fully overlapping trajectories indicate convergence for $dt < 0.2$. In \ta{},\tb{}, the lines are a guide to the eye, whereas in \tc{}-\tf{} the lines represent additional data points. \tc{},\td{} Phase-space evolution of a single molecule. Same color-code as in \ta{},\tb{}. \te{},\tf{} Time evolution of the right tail weight $\eta^r$ (see paper for definition). Same color-code as in \ta{},\tb{}. Parameters for all plots: $N=100$, $\lambda=0.5$, $\nu=0.3g_c$, $W=g_c/2$, $\chi=128$, $n_\mathrm{max}^v = 10$. Results for a single disorder realization.}
    \label{fig:dtconv}
\end{figure}

\begin{figure}
    \centering
    \includegraphics[width=0.5\textwidth]{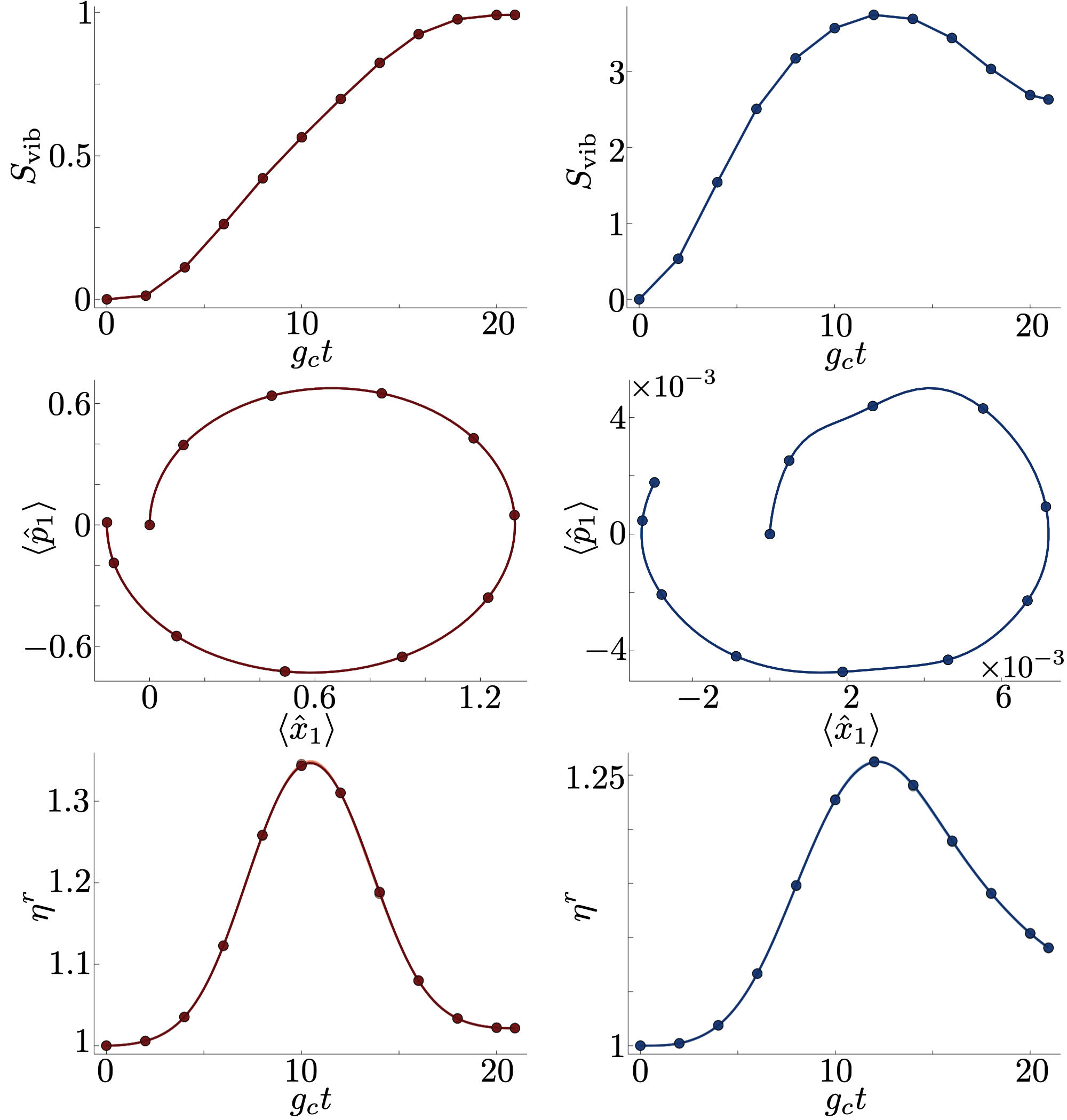}
    \caption{\textbf{Convergence with the maximum number of vibrations $n_\mathrm{max}^v$.} \ta{},\tb{} Time evolution of the vibrational entropy $S_\mathrm{vib}$ for an initial molecular (left, red) or cavity (right, blue) excitation. Darkness indicates $n_\mathrm{max}^v$. ($n_\mathrm{max}^v \in \{6,8,10,12\}$). Overlapping of the trajectories indicates convergence for $n_\mathrm{max}^v > 6$. In \ta{},\tb{}, the lines are a guide to the eye, whereas in \tc{}-\tf{} the lines represent additional data points. \tc{},\td{} Phase-space evolution of a single molecule. Same color-code as in \ta{},\tb{}. \te{},\tf{} Time evolution of the right tail weight $\eta_r$ (see paper for definition). Same color-code as in \ta{},\tb{}. Parameters for all plots: $N=100$, $\lambda=0.5$, $\nu=0.3g_c$, $W=g_c/2$, $\chi=128$, $dt=0.01$. Results for a single disorder realization.}
    \label{fig:nmaxvconv}
\end{figure}

Fig.~\ref{fig:chiconv} shows the time evolution of $S_\mathrm{vib}$, the phase space evolution, and $\eta^r$ computed with increasing bond dimension $\chi$ for a single disorder realization. We find that the vibrational entanglement $S_\mathrm{vib}$ generally reaches larger values for larger $\chi < 128$, and for $\chi = 128, 256, 512$ the data points overlap perfectly. This indicates convergence for $\chi=128$. For both the phase space evolution and the right tail the trajectories overlap for $\chi>16$ ($\chi>8$ for a molecular excitation), indicating much faster convergence for these observables. We attribute the slow convergence of $S_\mathrm{vib}$ with $\chi$ to the large number of swap operations $\sim N^2$ and the unfavorable MPS structure when computing $S_\mathrm{vib}$.

Fig.~\ref{fig:dtconv} shows the time evolution of the same observables computed using different time steps $dt$ used in the second order sweep. All lines overlap, indicating that convergence is already reached for $dt=0.2$. The small differences in the final data points are rounding errors $\sim dt$ for the choice of final time.

Fig.~\ref{fig:nmaxvconv} shows the time evolution of the same observables computed with different cutoffs for the number of vibrational excitations per molecule $n_\mathrm{max}^v$. We find again that all lines overlap, indicating convergence for $n_\mathrm{max}^v>6$.

\end{widetext}

\end{document}